\documentclass[letterpaper,twocolumn,10pt]{article}
\usepackage{usenix-2020-09}

\usepackage{tikz}
\usepackage{amsmath}
\usepackage{amsfonts}
\usepackage{amssymb}
\usepackage{bm}
\usepackage{tabularray}
\usepackage{algorithm}
\usepackage{algpseudocode}
\usepackage{algorithmicx}
\usepackage[capitalise,noabbrev]{cleveref} 
\Crefname{appsec}{appendix}{appendices}
\usepackage{enumitem}
\usepackage{graphicx}
\usepackage{subcaption}
\usepackage[normalem]{ulem}
\usepackage{xspace}
\usepackage{appendix}
\usepackage{url}

\newcommand{\name}{\textit{OverHear}\xspace}
\begin{document}

\date{}

\title{\Large \bf OverHear: Headphone based Multi-sensor Keystroke Inference}

\author{
{\rm Raveen Wijewickrama}$^{\ast}$,
{\rm Maryam Abbasihafshejani}$^{\ast}$,
{\rm Anindya Maiti}$^{\dagger}$,
{\rm Murtuza Jadliwala}$^{\ast}$ \\[2mm]
{$^{\ast}$University of Texas at San Antonio}\\
{\{raveen.wijewickrama, maryam.abbasihafshejani, murtuza.jadliwala\}@utsa.edu}\\
{$^{\dagger}$University of Oklahoma}\\
{am@ou.edu}
} 

\maketitle

\begin{abstract}
Headphones, traditionally limited to audio playback, have evolved to integrate sensors like high-definition microphones and accelerometers. While these advancements enhance user experience, they also introduce potential eavesdropping vulnerabilities, with keystroke inference being our concern in this work. To validate this threat, we developed \name, a keystroke inference framework that leverages both acoustic and accelerometer data from headphones. The accelerometer data, while not sufficiently detailed for individual keystroke identification, aids in clustering key presses by hand position. Concurrently, the acoustic data undergoes analysis to extract Mel Frequency Cepstral Coefficients (MFCC), aiding in distinguishing between different keystrokes.
These features feed into machine learning models for keystroke prediction, with results further refined via dictionary-based word prediction methods. In our experimental setup, we tested various keyboard types under different environmental conditions. We were able to achieve top-5 key prediction accuracy of around 80\% for mechanical keyboards and around 60\% for membrane keyboards with top-100 word prediction accuracies over 70\% for all keyboard types.
The results highlight the effectiveness and limitations of our approach in the context of real-world scenarios.
\end{abstract}

\vspace{-0.2cm}
\section{Introduction}
\label{sec:introduction}
\vspace{-0.35cm}
In the rapidly evolving landscape of mobile device hardware, sensors have emerged as pivotal components that enhance user experience and functionality. Among these sensors, accelerometers stand out for their ability to detect motion, while microphones, especially those embedded in headphones and headsets, have been refined to high-definition standards suitable for applications such as high fidelity voice recording and enabling active noise cancellation. Interestingly, these headphones not only come equipped with multiple microphones but also often incorporate accelerometers to discern if they are being worn, and for advanced applications such as gesture controls \cite{fan2021headfi,chen2015earpieces}, adaptive noise cancellation \cite{terlizzi2012systems}, spatial audio \cite{mao2012method}, and fitness tracking \cite{roddiger2022sensing,prest2014sports}. However, the integration of such advanced sensors presents potential security challenges. 
Given this rich sensor integration, headphones present themselves as an attractive and novel attack vector, especially for eavesdropping applications such as keystroke inference. Their proximity to the user and their surrounding devices (e.g., keyboards), combined with their multi-sensor capabilities, offers a unique advantage for capturing sensitive data. 
Despite these inherent advantages, the potential of headphones as a medium for keystroke inference has remained largely unexplored.

Acoustic eavesdropping has been a thoroughly investigated side-channel for inferring keystrokes on both physical and touchscreen keyboards. Lie et al. \cite{liu2015snooping} harnessed sound waves produced by typing on a traditional keyboard using a smartphone's microphone. In contrast, Narain et al. \cite{narain2014single} and Lu et al.'s KeyListener \cite{lu2019keylistener} focused on touchscreen keyboards, with the the latter using an adjacent adversarial phone. Similarly, Bai et al. \cite{bai2021know} applied stereo audio recording from a mobile phone near a physical keyboard for keystroke inference. These prior works mainly utilized fixed audio recording devices, such as mobile phones controlled by attackers or specific computer microphones, to detect the acoustic signals from keystrokes. In practical situations, it's unlikely that an individual would overlook unknown devices in their vicinity. Though an intruder could consider leveraging the victim's mobile phone (via a malicious app), the unpredictable and user-dependent placement of the phone relative to a keyboard often diminishes the feasibility of acoustic-only inference.

Past research has also explored the use of motion sensors for keystroke inference, often utilizing mobile phones or wearables such as smartwatches \cite{cai2012practicality, wang2015mole, liu2015good, maiti2016smartwatch}. Marquardt et al. \cite{marquardt2011sp} notably employed a smartphone's accelerometer to detect vibrations from nearby keyboard keystrokes. However, such methods are limited by the need for close and/or fixed proximity to the keyboard or require the victim to wear or hold the infiltrated wearable device on their hand, limiting their practicality (as in the earlier case).

Diverging from the traditional attacker controlled devices or infiltrated user mobile phone based methods, we look into the possibility of using smart headphones for keystroke inference. Modern headphones, especially the latest Bluetooth-enabled variants, often come equipped with multiple microphones on both ears. These microphones serve dual purposes: capturing audio and facilitating noise cancellation. Such a design makes headphones an enticing vector for a keystroke inference attack. Further, recognizing that the latest varieties of headphones, particularly Bluetooth enabled ones, come with smartphone/desktop apps for their respective devices, we speculate that a malicious app which has firmware level API access to the headphone can potentially exploit this connection to discreetly record sensor data in order to carry out keystroke inference on unsuspecting users. A likely scenario would be an instance where a manufacturer itself acting as an adversary using its native app to offload data to its back-end server in order to further process this data to infer sensitive information about the users beyond the recorded audio.

One of the significant challenges in the domain of keystroke inference is training-free inference, where an attacker lacks labeled training data specific to the victim. More specifically, in a headphone-based keystroke inference attack scenario, the relative position between the headphones and the keyboard can vary considerably based on user traits such as height, arm length, and the distance from the chair to the table. This is compounded by both voluntary and involuntary head movements. Prior research that utilized dual microphones, primarily relied on a technique called Time Difference of Arrival (TDoA) which uses time difference that results when sound travels to the individual microphones, yielding stable results primarily because the data collection device was either on the same surface as the keyboard or was the input device itself (e.g., mobile phone qwerty keypad keystroke inference) \cite{yu2021security}. 
Further, these prior works on acoustic keystroke inference relied on recording devices that were mostly stationary, while in our situation, due to the motion of the headphone induced by user head movements, the recording devices are not static and thus may record much more noisier/inconsistent data.

Given the opportunity of being able to employ modern headphones equipped with a variety of on-board sensors as a novel attack vector and the associated challenges, in this paper we propose \name, a framework to infer keystroke using acoustic and motion sensor data collected from headphones. While the accelerometer data alone lacks the granularity to distinguish individual keystrokes, it proves to be useful in clustering keys corresponding to each hand especially when traditional acoustic based sound source localization techniques may not be working well due to potentially unsteady behavior of the victims.
From the acoustic data, we extract Mel-frequency cepstral coefficients (MFCC) features, and train and test several machine learning models to predict individual keys. This prediction is then refined using a dictionary-based spell-correction approach to further improve the success of a keystroke inference in a context-aware manner. 
We attained a top-5 key prediction accuracy of approximately 80\% for mechanical keyboards and about 60\% for membrane keyboards. Furthermore, our framework demonstrated a top-50 word prediction accuracy nearing 50\%, and surpassed 70\% in top-100 word prediction accuracy across all keyboard types.

Our main contributions are as follows. 
\begin{itemize}[leftmargin=*]
\setlength{\itemsep}{0pt}
  \item \textbf{Development of a new keystroke inference framework}: To overcome the unique challenges encountered in headphone based keystroke inference, we propose a novel inference framework called \name. \name integrates data from both microphones and accelerometers to enhance the accuracy of keystroke inference. \name also includes a keyboard type identification module to identify the type of the keyboard a victim may be using (e.g., mechanical or membrane).

  \item \textbf{Enhanced word prediction mechanism}: We incorporate a word prediction technique based on spell-correction to further improve the efficacy and prediction performance of \name.
  
  \item \textbf{Comprehensive empirical evaluation of the proposed attack}: \name is evaluated using data sourced from real-world participants under realistic/unconstrained settings, spanning across various environmental/ambient noise scenarios. This ensures that the findings are reflective of practical, day-to-day scenarios, offering insights into the framework's robustness and applicability.
\end{itemize}

\vspace{-0.2cm}
\section{Motivation and Related Work} %
\label{sec:related_works}
\vspace{-0.35cm}
Analysis of acoustic signals for the purpose of inferring user input on a variety of mobile and computing devices has been the subject of several studies in recent years, each employing unique methodologies and achieving varying degrees of accuracy.
Asonov and Agarwal \cite{asonov2004keyboard} in 2004 employed acoustic data collected using a dedicated PC microphone with a trained neural network that uses frequency domain features to show that individual key presses can be recognized with an accuracy of 79\%. They demonstrated that their inference framework works across multiple different computer keyboards as well as telephone and ATM key pads. 
In a similar line of research, Liu et al. \cite{liu2015snooping} proposed a novel approach for inferring keystrokes on a mobile device by using the resulting audio signals. They developed a system that utilizes the built-in stereo microphones of a smartphone positioned adjacent to keyboard to record sound waves produced by a user's typing. By means of a purely non-ML based approach such as a Time-Difference-of-Arrival (TDoA) technique, they were able to identify unique patterns in these recorded acoustic waves and match them to specific keys (being typed). Their framework was able to produce key inference accuracies close to 85\%.
While both the above two research efforts are significant, they were tested under rather restrictive conditions and focused solely on individual key presses rather than more complex scenarios involving word or sentence typing and prediction. 

Similar to Asonov and Agarwal \cite{asonov2004keyboard}, Zhuang et al. used a Hidden Markov Model (HMM) to infer keystrokes from audio signals recorded by a PC microphone, and then employed a language model to facilitate word prediction. They were able to achieve a prediction accuracy of 88\% and 96\% for word and keys, respectively. 
Narain et al. \cite{narain2014single} proposed a framework to infer keystrokes on a touchscreen QWERTY keyboard of a mobile device using acoustic signals. 
They employed a Decision Tree based learning algorithm, achieving a fairly high accuracy of close to 95\% in a single attempt. The most notable aspect of this study was its language-agnostic approach, suggesting that regardless of the language or the content being typed, the methodology can accurately infer the keystrokes.

Similar to Narain et al.'s approach \cite{narain2014single}, Lu et al.'s KeyListener \cite{lu2019keylistener} attempts to infer touch screen QWERTY keyboard keystrokes on a smartphone, but with a nearby adversarial smartphone's microphone. 
They used a Time-Difference-of-Arrival (TDoA) based approach, achieving a top-1 word accuracy of around 50\% and a  top-10 accuracy of around 90\%.
Zhu et al. \cite{zhuacoustic2014} also proposed keystroke inference framework based on TDoA and achieved a key prediction accuracy of 72\%. However, their framework requires at least two malicious mobile phones to be in physical proximity of the victim keyboard which making executing the attack practicality challenging.
Lastly, Bai et al. \cite{bai2021know}, drawing parallels with Lie et al. \cite{liu2015snooping} and Zhu et al. \cite{zhuacoustic2014}, employed stereo audio recordings from a mobile phone situated near a keyboard to deduce keystrokes. Their methodology integrates (TDoA) and Power Spectral Density (PSD) features within a SVM model. They achieved a top-1 accuracy of 71\% and a further enhanced top-5 accuracy of 92\%. 

Several past studies have also explored the use of motion sensors for keystroke inference \cite{cai2012practicality, wang2015mole, liu2015good, maiti2016smartwatch}. These investigations typically harness motion sensors in mobile phones to deduce in-device keystrokes or employ smart wearables, like smartwatches, to infer keystrokes on physical keyboards. Marquardt et al. \cite{marquardt2011sp} introduced (sp)iPhone, where accelerometer data from a device placed near a physical keyboard could be used to infer keystrokes. For their attack, they capitalized on the vibrations generated by the keystrokes on the keyboard and transmitted through the table to the smartphone's accelerometer. 

Our extensive literature review has shown key research gaps, which motivates us to explore the feasibility of a new attack surface for carrying out keystroke inference attacks (on external keyboards). First, previous research which employed acoustic and/or motion/vibration signals for such attacks primarily relied on stationary recording sources/devices, such as attacker-controlled mobile phones or dedicated PC microphones, to capture the motion/vibration and/or sounds produced by the keystrokes. In real-world scenarios, it's unlikely that a victim wouldn't notice unfamiliar (recording) devices nearby which makes actually carrying out such attacks challenging. While an attacker might attempt to use the victim's own mobile phone (by installing some Trojan app) for such purposes, the unpredictable positioning of the phone near a keyboard makes this approach less practical. Further, headphones, especially the wireless type, are ubiquitous and often worn continuously by users, even when not in active use  (e.g., for noise cancelling purposes). This constant presence provides a persistent eavesdropping opportunity.

Our attack model diverges from these existing approaches by leveraging the user's own headphones, which typically has a fixed position over the user's head. Given that many of these headphones come with companion apps for smartphones or desktops (especially modern Bluetooth ones), we assume that a malicious companion app could serve as a side-channel to extract sensor data (e.g., acoustic and motion sensors) from the unsuspecting victim in order to carry out keystroke inference attacks. However, our attack setup is not free of challenges. The acoustic stream in our attack scenario presents its own set of challenges. Unlike stationary microphones or mobile devices placed on a surface, headphones are subject to the nuances of human behavior. The user's head movements, whether subtle shifts or more pronounced turns (while typing), can introduce variability in the captured acoustic data. Furthermore, individual anatomical traits, such as the shape and size of the user's head, length of the arms, can influence how sound waves are received by the headphone microphones.
On the other hand, the motion sensor data streams have their own challenges, i.e., the vibrations (due to keystrokes) sensed by the on-board motion sensor are less pronounced as they have to traverse from the hands/body to the head of the user. Consequently, relying solely on accelerometer or motion sensor data for keystroke inference would not be very effective. Our key insight is whether acoustic/sound data as sensed by the microphones can be combined with motion sensor data, both of which are readily available in most modern headphones, can be used to effectively carry out keystroke inference in this unique, yet realistic, setup. We are the first to explore this novel attack setup for keystroke inference.

\vspace{-0.2cm}
\section{Background and Preliminaries}
\label{sec:background}
\vspace{-0.35cm}
This section briefly introduce the type of sensors used in our keystroke inference framework (\name ), their sensor feedback, and the concept of Time Difference of Arrival (TDoA) which are required to understand the rest of the paper.

\vspace{-0.2cm}
\subsection{Microphones in Headphones} %
\label{sub:acoustic_signal}
\vspace{-0.2cm}
Microphones function by capturing sound waves (acoustic energy) and converting them into electrical signals that can then be recorded and utilized in various audio and voice applications.
The electrical signal is generated by vibrations of a diaphragm or membrane, which moves a coil of wire or a capacitor in response to sound waves. Microphones come in a variety of forms, including dynamic, ribbon, condenser, MEMS (Micro-Electro-Mechanical Systems) and electret.
The two most common polar patterns used in microphones are omnidirectional and cardioid. In an omnidirectional microphone, sound is picked up equally from all directions. As a result, it can detect sound coming from the front, sides, and back of the microphone. Conversely, a cardioid microphone is unidirectional, which means it picks up sound primarily from one direction (usually the front) and rejects it from other directions. In modern headsets and earbuds where the microphones are built into the earpiece itself (compared to the ones where a microphone arm protrudes from the earpiece to near the mouth), omnidirectional microphones are used. This is mainly due to the positioning, since the microphones are not placed near the mouth, it would be challenging for the a cardioid microphone to consistently pick up user voice. However, one drawback of using omnidirectional microphones is the potential for more background noise. To overcome this, many modern high-end headsets employ noise-cancellation techniques to help isolate user's voice and reduce ambient noise primarily aided by the presence of multiple microphones present in these headphones, i.e., for user input and dedicated noise cancelling purposes\cite{liebich2018signal,sank1984microphones}.

\vspace{-0.2cm}
\subsection{Motion Sensors in Headphones} %
\vspace{-0.2cm}
There has been a rising trend in incorporating intelligent features into headphone, in order to enable smart applications. Early adapters of this innovation, such as the Microsoft Surface headphones and Bose QC35 headphones, have embedded motion sensors, enhancing user interaction with gesture-based controls. These headphones empower users to perform actions like playing or pausing audio and summoning voice assistants like Siri or Alexa through intuitive gestures\cite{fan2021headfi,chen2015earpieces}.
Central to these advancements are two critical sensors: accelerometers and gyroscopes. Together, these sensors capture movements of the device in three-dimensional space, breaking them down along the $x$, $y$, and $z$ axes. While accelerometers focus on linear acceleration -- encompassing actions such as tilting or straightforward motion -- the gyroscopes are designed to measure angular velocity, effectively capturing rotational movements around the three principal axes. This ability allows for precise discernment of the device's orientation and angular shifts\cite{xu2012taplogger}.

\vspace{-0.2cm}
\subsection{Sensor Fusion for Keystroke Inference}
\begin{figure}[t]
    \centering
    \begin{subfigure}{0.49\linewidth}
        \centering
        \includegraphics[width=\linewidth]{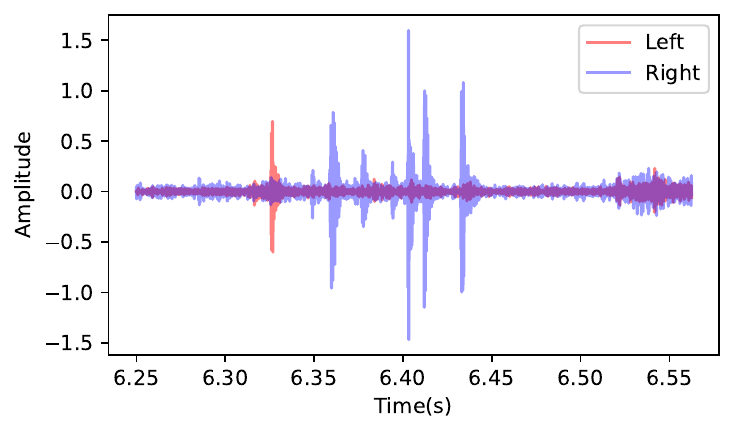}
        \caption{}
        \label{fig:audio_sample}
    \end{subfigure}
    \hfill
    \begin{subfigure}{0.49\linewidth}
        \centering
        \includegraphics[width=\linewidth]{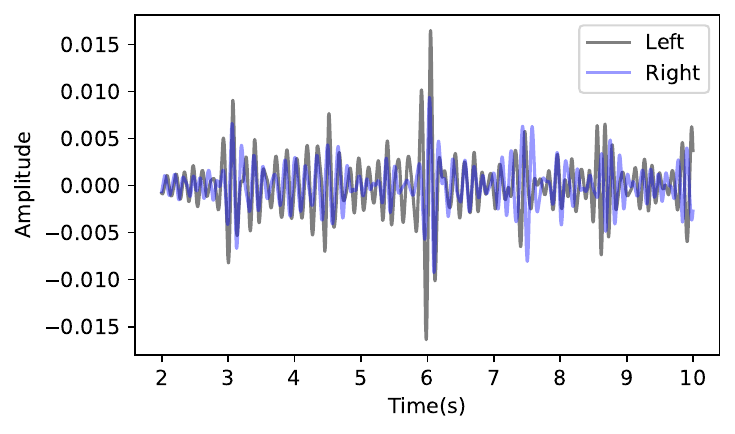}
        \caption{}
        \label{fig:acc_sample}
    \end{subfigure}
    \vspace{-0.375cm}
    \caption{Typing captured through a pair headphones, (a) audio, (b) accelerometer.}
    \label{fig:audio_acc_samples}
    \vspace{-0.25cm}
\end{figure}

\vspace{-0.2cm}
Our particular interest is in the potential application of these sensors beyond their intended use, specifically in keystroke inference. Keyboards possess distinct mechanical characteristics, resulting in the emission of unique acoustic signatures when keys are pressed and released. As leveraged in previous works~\cite{bai2021know,asonov2004keyboard}, these audible vibrations are detectable by microphones integrated into nearby devices (see \cref{fig:audio_sample}). Furthermore, as a user engages with a keyboard, the act of typing creates additional non-acoustic vibrations. These vibrations, originating from keystrokes, could travel through the fingers, palms, and further along the arms. \Cref{fig:acc_sample} shows the accelerometer feedback recorded from a pair of prototype headphones (equipped with accelerometers on each ear piece) during a typing task. The noticeable peaks correlate to the key presses that are occurring indicating the potential of accelerometers for keystroke inference. Though the strength of these waves diminishes with distance, the sensitivity of motion sensors in headphones might be adept enough to pick up certain subtle motions which could facilitate keystroke inference. This ability to capture such nuanced data from motion sensors by combining it with audio data could be instrumental in building a more comprehensive detection and inference system. Such data fusion provides a comprehensive view of the data, capturing different perspectives and nuances. As a result of the integration of these diverse data streams, the detection system overcomes the limitations and biases inherent in individual sensors, enhancing the identification of patterns. Additionally, integrating multiple data reduces false positives and false negatives, resulting in better accuracy.

\vspace{-0.2cm}
\subsection{Sound Source Localization via Time Difference of Arrival (TDoA)} %
\label{sub:time_difference_of_arrival_}
\vspace{-0.2cm}
Time Difference of Arrival (TDoA) is a method used in acoustic, radar, or radio signals to estimate the location of an object using two or more sensors. 
The overarching principle behind TDoA is the sound produced by a source travels through a medium (often air) and reaches one microphone before the other, if the sound source isn't equidistant to the two microphones. This difference in time that the sound signal takes to reach each microphone is appositely called the Time Difference of Arrival. 
Although the process of calculating TDoA is simple in theory, it becomes complex in practice due to factors like signal distortion, background noise, and reflection of sound waves from surfaces. These factors can affect the signal and make it more difficult to accurately identify the arrival time of the signal at each microphone. Several acoustic based inference works in the past have successfully used TDoA based techniques to predict keystroke \cite{liu2015snooping,cheng2022dictionary}.
However, the problem becomes even more challenging if the microphones are not in a fixed position, such as from a pair of headphones due to the head movements that may occur while a person is wearing the headphones. Further, specifically in an inference attack scenario, victim specific anatomical differences (e.g., height, arm length) and voluntary/involuntary head movements could make the use of TDoA less effective. 
In \cref{sub:clustering} we detail the observations we made using real-world participants data on shortcomings of TDoA in our attack scenario and describe the alternative techniques we used to overcome such challenges.

\vspace{-0.2cm}
\section{\name Overview} %
\label{sec:adversary_model}
\vspace{-0.35cm}
In this section, we describe our adversary model followed by an overview of our inference attack.

\vspace{-0.2cm}
\subsection{Adversary Model} %
\label{sub:adversary_model}
\vspace{-0.2cm}
The primary objective of an adversary in our attack is to infer sensitive information typed by the user on a physical keyboard, such as passwords, credit card numbers, and personal identification details.
We assume that the adversary has the ability to access and acquire both audio and accelerometer data from the target user’s headphones where the adversary intercepts the communication between the headset and the connected device. This can be done via either a Trojan application installed in the paired device such as a desktop computer/laptop or a mobile phone which will have firmware level API access to the headphones. 
Such an attack setup or adversarial scenario can be easily realized by the headphone device manufacturer themselves acting as the adversary \cite{anker_cam,ring_bbc,vizio_spy}.
The captured data is then transferred to an adversary controlled server elsewhere in which the victim data will be run through a pre-trained model to infer/predict what the target victim typed.
The adversary has knowledge of the type of keyboard (i.e., a mechanical keyboard, membrane keyboard (external) or a laptop membrane keyboard) the victim uses by employing a keyboard identification model as described in \cref{sub:keyboard_type_inference}.
The adversary does not have any other medium of inferring the private text typed, and must rely entirely on the data streams originating from the headphones.

\vspace{-0.2cm}
\subsection{Inference Framework} %
\begin{figure}[t]
\centering
\includegraphics[width=0.99\linewidth]{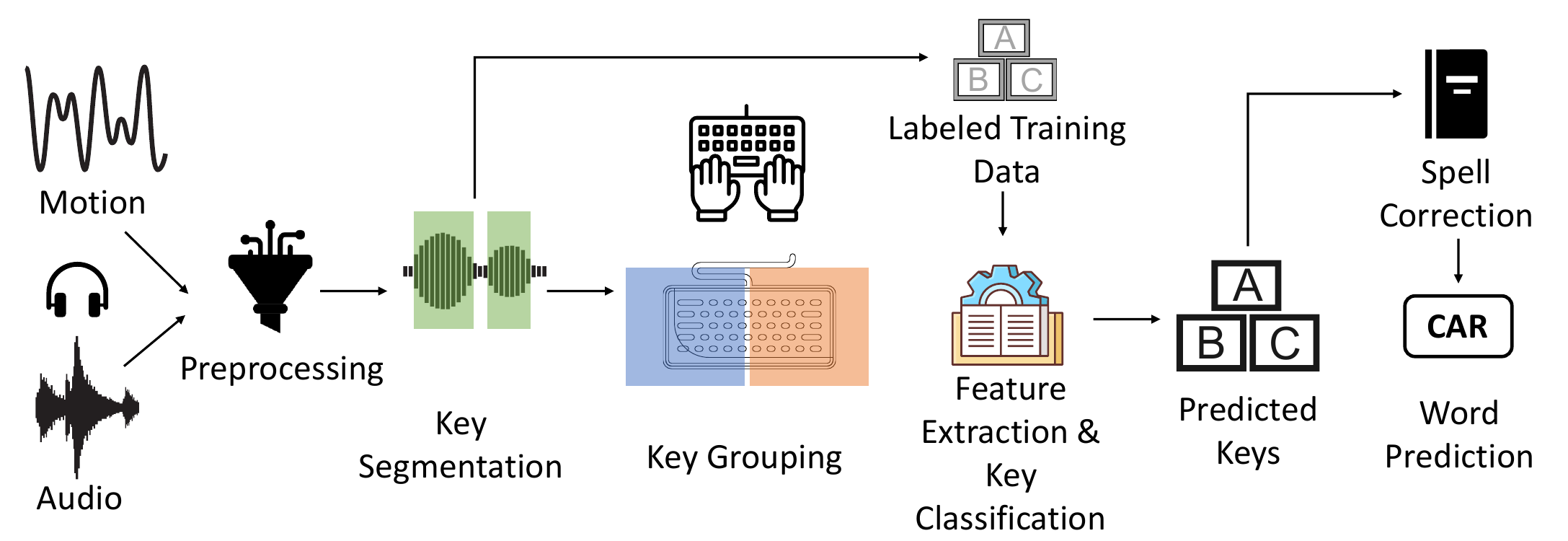}
\vspace{-0.2cm} %
\caption{\name inference framework overview.}
\label{fig:attack_framework}
\vspace{-0.25cm}
\end{figure}

\label{sub:attack_overview}
\vspace{-0.2cm}
\Cref{fig:attack_framework} illustrates our \name inference framework architecture. A malicious companion app associated with the victim's headphones is used to covertly capture data from the headphones' built-in accelerometers and stereo microphones during typing activities. 
For building the training dataset, a custom application connected to the headphones are used to capture the data streams.
The raw accelerometer and audio data are then transmitted to a remote server, which houses the remaining components of our inference framework. This server processes the data, constructs training models, and conducts evaluations on the test data.

\noindent
\textbf{Recording and Data Pre-processing.} \name uses raw audio and accelerometer data captured through their headphones during keyboard interactions of the victim. The type of keyboard in use is then identified to tailor the subsequent processing steps. Noise filtering techniques are applied to remove any background/ambient noise that may be present in the recorded streams. Once cleaned, a segmentation algorithm is employed to isolate and identify individual keystrokes.

\noindent
\textbf{Training Dataset Assembly.} Using the insights gained from the pre-identified keyboard type and other criteria, a comprehensive training dataset is curated using acoustic and motion sensor data captured using headphones. This involves data collection sessions with a select group of participants, ensuring a diverse and representative sample that captures various typing patterns and styles.

\noindent
\textbf{Feature Extraction and Model Training.} With the training data in place, the system extracts a set of features for each keystroke which encapsulate the unique characteristics of keystrokes. The keys are then clustered into three groups based on their potential typing hand; left, right and ambiguous (see \cref{ssub:energy_based_clustering} for more details) and
then used to train a machine learning model, optimizing it for accuracy and generalization across different users/victims with varying typing traits.

\noindent
\textbf{Prediction on Victim Data.} Using the trained inference model from the previous step, the framework now processes unlabeled data from the victim to infer their keystrokes. The audio and motion sensor data streams captured through victim's headphones go through the same pipeline as the training dataset where the data streams are first pre-processed and segmented to identify keystrokes. The unlabeled keystrokes are then tested via the trained machine learning models to predict keystrokes. The sequences of keystrokes are then further processed in a word prediction module with the aid of a spell correcting algorithm to predict the closest matching words.

\vspace{-0.2cm}
\section{Framework Design and Implementation} %
\label{sec:attack}
\vspace{-0.35cm}
In this section, we outline the design of our \name framework and then discuss the specifics of its implementation, including our experimental setup and data collection procedure.

\vspace{-0.2cm}
\subsection{Data Pre-processing}
\label{sub:preprocessing}
\vspace{-0.2cm}
\noindent
\textbf{Audio Noise Filtering.}
To ensure the clarity and relevance of our audio data in the context of keystrokes, we employed filtering to mitigate background noise, which typically occurs at higher frequencies distinct from those of keystrokes. Through analysis of the audio captured via our headphone setup, we discerned that keystrokes predominantly occur within the frequency range of 1200 - 3800 Hz. Consequently, a bandpass filter was tailored to retain information within this specific range while filtering out extraneous frequencies. However, while this range effectively captures both mechanical and membrane type keyboard keystrokes tested in our study, adjustments might be necessary to accommodate the unique sound profiles of other keyboards.

\noindent
\textbf{Accelerometer Noise Filtering.}
The raw accelerometer data very often contain noise, primarily stemming from involuntary body movements, if from a body worn device. Specifically, in our headphone based attack scenario, this can be accounted to minor head movements. To address this, we employed a low-pass filter designed to eliminate high-frequency noise while preserving the lower-frequency vibrations induced by key presses. 

\vspace{-0.2cm}
\subsection{Keystroke Segmentation}
\label{sub:segmentation}
\begin{figure}[h]
\centering
\includegraphics[width=0.7\linewidth]{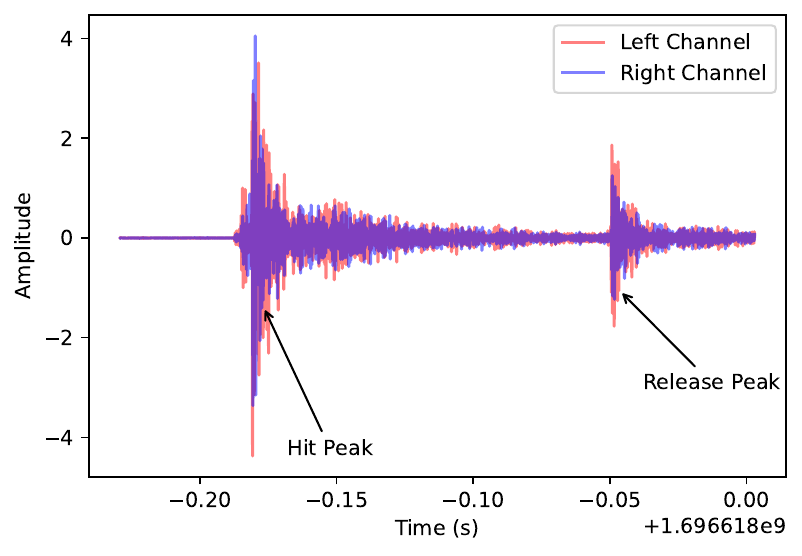}
\vspace{-0.375cm}
\caption{Acoustic waveform of a keystroke.}
\label{fig:audio_keystroke}
\vspace{-0.25cm}
\end{figure}

\vspace{-0.2cm}
A keystroke consists of two main components in its acoustic signal, namely the the key hit and key release which produces two different peaks in the signal.
\Cref{fig:audio_keystroke} illustrates the typical acoustic feedback captured from the stereo microphones on a pair of headphones from a keyboard key press event. The initial more pronounced peak represents the key hit event, while the subsequent, lower peak indicates the key release event. 

\vspace{-0.25cm}

We observed that the average duration of a keystroke is about 80~ms. We use a sliding window with a size of 1~ms (empirically determined) and calculates the energy (\cref{eq:1_energy}), which is the sum of the squares of the audio amplitude values for each window and is given by the following formula:
\begin{equation}
\label{eq:1_energy}
E_x = \sum_n |x(n)|^2
\end{equation}
where $E$ is the energy of the signal, $n$ is the index of the audio sample and $x(n)^2$ is the square of the amplitude of the signal at the $n^{th}$ sample.
We then pass each window through an adaptive thresholding peak picking algorithm called \emph{Music Structure Analysis Framework (MSAF)} \cite{nieto2015msaf} which considers a local average to pick the prominent keystroke related peaks and to discard insignificant noise peaks. The windows with the detected peaks are then considered to be potential keystroke start points, $p_s$. The consecutive start points less than 100~ms are discarded. For each start point, we extract the keystroke as follows:
\vspace{-0.2cm}
\begin{equation}
ks_i = (p_{si}-5\text{{ms}}, p_{si}+80\text{{ms}})
\end{equation}
where, $ks_i$ is the $i^{th}$ keystroke in the continuous signal, $p_{si}$ is the start point of the keystroke, and $p_{si}+80ms$ is the end point of the keystroke.

\vspace{-0.2cm}
\subsection{Key Group Clustering}
\label{sub:clustering}
\vspace{-0.2cm}
One of the challenges in a keystroke inference attack is the number of potential keys on a keyboard, which can make the prediction task complex. Training a single model to distinguish between each key individually may require a vast amount of data for each key to achieve reasonable accuracy. By grouping keys, the dimensionality of the problem can be reduced, making the training and inference processes more practical. 
We first look into methods used for key grouping in previous acoustic based keystroke inference works \cite{narain2014single,liu2015snooping} for their applicability in our attack scenario, identify challenges posed by them, followed by proposing techniques that suits our headphone based inference setting.

\vspace{-0.2cm}
\subsubsection{Traditional TDoA Based Clustering} %
\label{ssub:tdoa_based_clustering}
\vspace{-0.2cm}
We first look into the possibility of using traditional Time Difference of Arrival (TDoA) based key identification/key group clustering methods to identify similar keystrokes with similar sound profiles\cite{cheng2022dictionary}.
We compute TDoA via the cross-correlation method for the 2-channel audio signal. TDoA estimation using cross-correlation involves finding the lag (or shift) at which two signals are most similar \cite{bai2021know}. 
The cross-correlation TDoA formula is as follows:
\begin{equation}
\text{{TDoA}} = \arg\max_k \left( \sum_n S_1[n] \cdot S_2[n+k] \right)
\end{equation}
where \( S_1[n] \) is the signal at the first microphone at discrete time \( n \), \( S_2[n+k] \) is the signal at the second microphone shifted by \( k \) samples and \( \arg\max_k \) indicates the shift \( k \) at which the cross-correlation is maximized, i.e., the shift where the two signals look the most similar.

\begin{figure}[h]
    \centering
    \begin{subfigure}{0.49\linewidth}
        \centering
        \includegraphics[width=\linewidth]{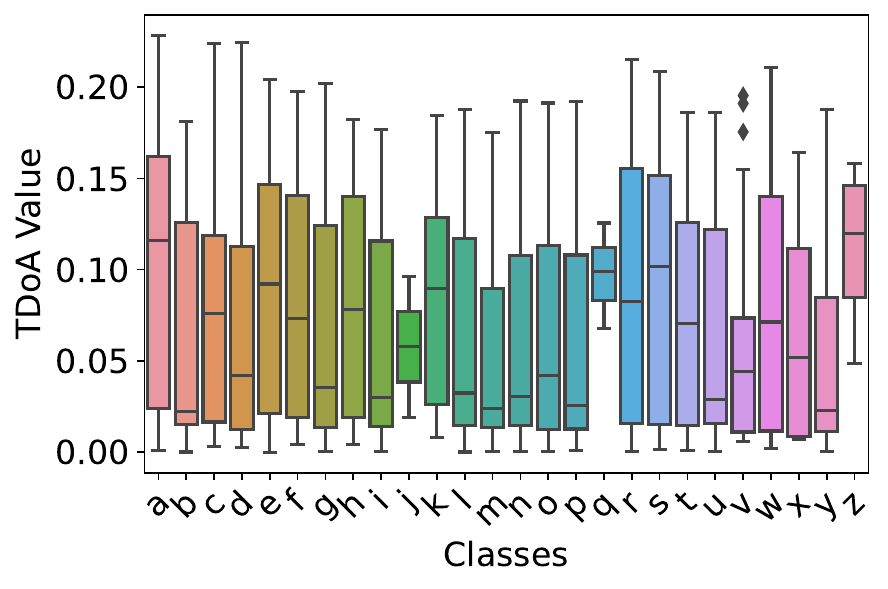}
        \caption{}
        \label{fig:tdoa_var_single}
    \end{subfigure}
    \hfill
    \begin{subfigure}{0.49\linewidth}
        \centering
        \includegraphics[width=\linewidth]{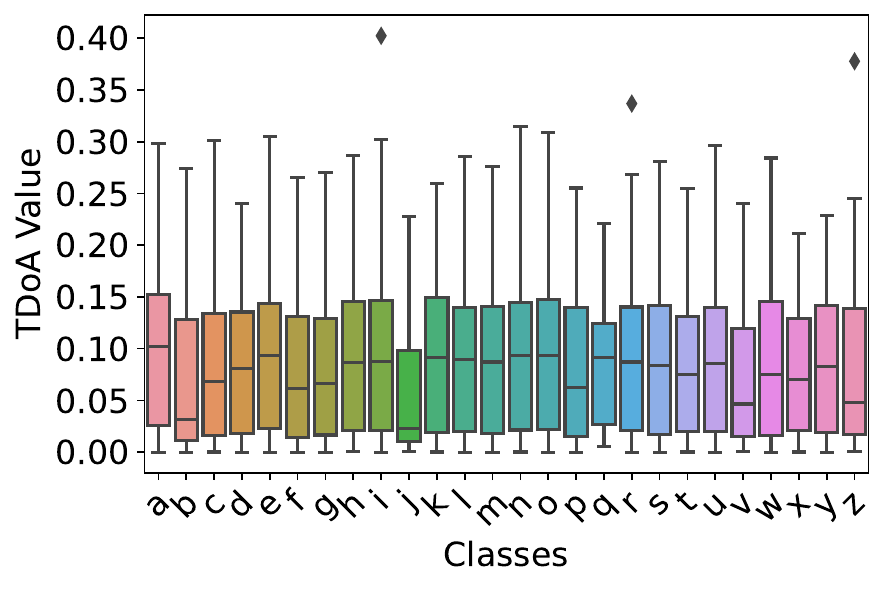}
        \caption{}
        \label{fig:tdoa_var_all}
    \end{subfigure}
    \vspace{-0.375cm}
    \caption{Variability of TDoA values (a) for a single participant, (b) across all participants.}
    \label{fig:tdoa_variability}
    \vspace{-0.25cm}
\end{figure}

However, the dynamic nature of head movements during typing tasks presented a significant challenge. Typists frequently shift their gaze, alternating between various sections of the screen and sometimes the keyboard. This constant change in head orientation rendered TDoA an unreliable metric for uniquely determining key positions. 
This is evident not only for individual users/typists, as shown in \cref{fig:tdoa_var_single}, but also when considering multiple users, as depicted in \cref{fig:tdoa_var_all}. The extensive spread of data points across all keys or classes further corroborates this observation. 
Moreover, our attempts to identify distinct key groups/clusters based on similar TDoA values, as was done in some previous works \cite{liu2015snooping,cheng2022dictionary}, proved to be unsuccessful. The high variability and inconsistency in TDoA values, exacerbated by the previously mentioned anatomical differences and head movements of the users, rendered the task of grouping keys with similar acoustic characteristics nearly impossible. This further highlights the unique challenges posed by our headphone-based setup compared to previous keystroke inference methodologies \cite{liu2015snooping,narain2014single,bai2021know}.

\vspace{-0.2cm}
\subsubsection{Energy Based Clustering} %
\begin{figure}[h]
    \centering
    \begin{subfigure}{0.49\linewidth}
        \centering
        \includegraphics[width=\linewidth]{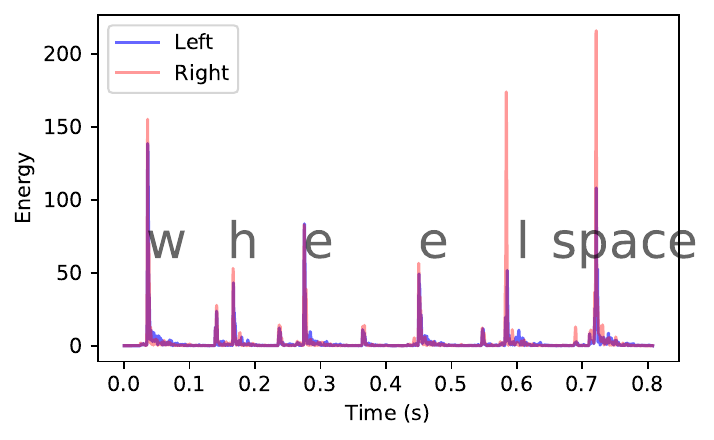}
        \caption{}
        \label{fig:audio_energy_diff}
    \end{subfigure}
    \hfill
    \begin{subfigure}{0.49\linewidth}
        \centering
        \includegraphics[width=\linewidth]{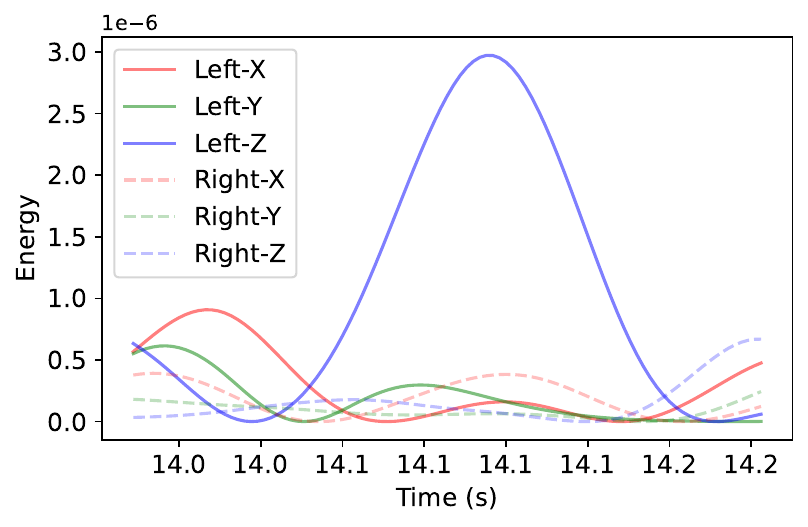}
        \caption{}
        \label{fig:acc_energy_diff}
    \end{subfigure}
    \vspace{-0.375cm}
    \caption{Difference between energy levels of left and right (a) audio channels for keystrokes when typing the word ``wheel'', (b) accelerometer channels when pressing key `a'.}
    \label{fig:aud_acc_energy_diff}
    \vspace{-0.25cm}
\end{figure}

\label{ssub:energy_based_clustering}
\vspace{-0.2cm}
We next explored other techniques that could aid our framework in identifying key groups. Specifically, we explored the energy levels of keystrokes in the acoustic signal.
As we can observe in \cref{fig:audio_energy_diff}, the energy for the keystrokes towards the right of the keyboard is higher on the right audio channel compared to left channel and vice versa. Under a setting similar to previous works \cite{liu2015snooping,bai2021know}, where the audio recording happens from a fixed position such as a phone kept nearby the keyboard, this type of energy based clustering can easily be used to cluster the keys into left and right groups. 
However, due to the constantly varying head direction changes that happen during typing tasks, which may include either looking at different parts of the screen or looking at the keyboard and then looking back at the screen, the energy differences for left and right audio channels also turned out to be not reliable and consistent.

\vspace{-0.25cm}

Although, the audio channels are affected by the head movements and deemed unreliable for the use of key clustering, we discovered an alternative in the accelerometers embedded on both sides of the headset. These accelerometers showed potential in estimating whether a keystroke was made by the right or left hand, based on the motion feedback they recorded.
Specifically, the upward and downward head movements, which often occur during typing (for instance, when participants glance at the keyboard and then refocus on the screen), is predominantly observed on the $x$-axis of the headphone accelerometers.
The side-way head movements that may occur during typing are noticeable on the $y$-axis. 
The $z$-axis was observed to be the most stable axis to capture the key press vibrations belonging to left or right hand. \cref{fig:acc_energy_diff} shows the accelerometer feedback captured from from our prototype (detailed in \cref{sub:experimental_setup}) headphone.
when pressing key `a' using the left hand in which the z axis energy profile of the left channel is clearly distinct from all the other axes.

With the aforementioned observations on energy significance on accelerometer z-axis, to quantify the distribution of energy between the left and right accelerometer channels, we introduce an energy ratio metric. For each channel, the energy, \( E \), is computed as the sum of the squares of its samples (see \cref{eq:1_energy}).
The energy ratio, \( E_R \), is then defined as the proportion of the left channel's energy to the total energy of both channels, formulated as:
\begin{align}
E_R &= \frac{E_{\text{left}}} {E_{\text{left}} + E_{\text{right}} + \epsilon}
\end{align}
where \( \epsilon \) is a minuscule constant introduced to prevent division by zero.
This metric provides a relative energy measurement between the two channels.
If a key is pressed from the left-hand, the energy registered on the left accelerometer channel (of the headphones) will be higher than the right accelerometer making the $E_R$ closer to 1 and if it's a right-hand pressed key, $E_R$ will be closer to 0. Since the energy ratio inherently normalizes the values between 0 and 1. This makes it easier to adapt across different users, as the absolute energy values can vary based on factors like distance from the source of vibration/key press (due to anatomical differences), or even the user's typing intensity.

Given the energy ratio, we heuristically %
label three key groups based on their spatial positioning and the hand predominantly used to press them. Specifically, we define:
\begin{itemize}[label={},leftmargin=*]
\setlength{\itemsep}{0pt}
    \item $\bm{G_1}$: Keys predominantly pressed by the left hand, namely \{a, s, d, z, x, q, w\}.
    \item $\bm{G_2}$: Keys predominantly pressed by the right hand, namely \{o, p, k, l, n, m, i, j\}.
    \item $\bm{G_3}$: Ambiguous keys that could be pressed by either hand, namely \{r, t, y, u, f, g, h, v, b, c, e\}.
\end{itemize}

We observed that keys within groups $G_1$ and $G_2$ are predominantly pressed by the left and right hands, respectively, across different users, especially due to their spatial positioning on the keyboard (extreme left and extreme right).
However, the keys in group $G_3$ presented ambiguity, with the choice of hand varying from one user to another. Such variations could arise from individual typing habits or a user's inclination to favor their dominant hand.
During the testing phase on unseen data, the median energy ratio, \(E^{med}_R\), is computed for all samples for a given test user/victim. The rationale behind computing the median energy ratio is to account for the variability in key pressing intensities among different participants. Different participants may exert different pressures when pressing keys, leading to variations in the vibrational feedback recorded by the accelerometers. By using the median, we aim to normalize this variability and achieve an adaptive clustering mechanism.
Keys with an energy ratio greater than $E^{med}_R$ are classified into $G_1$, while those with an energy ratio less than $E^{med}_R$ are classified into $G_2$. Keys with an energy ratio close to $E^{med}_R$, within a threshold $\gamma$, are classified as $G_3$.

Later in \cref{sub:feature_extraction} we train three different classification models, one for each group, to predict exact keys within each group.
During this prediction, if a test keystroke is initially classified into $G_1$ but the prediction probability is below a certain threshold $\lambda$, the test keystroke is then also evaluated by the models for $G_2$ and $G_3$. The final prediction is chosen from the model that yields the highest prediction probability (see \cref{alg:clustering}).

\vspace{-0.2cm}
\begin{algorithm}
\small
\caption{Energy Ratio-Based Key Group Classification}
\label{alg:clustering}
\begin{algorithmic}[1]
\Require \(E_R\), \(E^{med}_R\), \(\lambda\)
    \If{\(E_R\) > \(E^{med}_R\)}
        \If{Prediction Probability of \(G_1\) < \(\lambda\)}
            \State Evaluate using \(G_2\), \(G_3\)
            \State MaxProbability(\(G_1\),\(G_2\),\(G_3\))
        \Else \textbf{classify} using \(G_1\)
         \EndIf
    \ElsIf{\(E_R\) < \(E^{med}_R\)}
        \If{Prediction Probability of \(G_2\) < \(\lambda\)}
            \State Evaluate using \(G_1\), \(G_3\)
            \State MaxProbability(\(G_1\),\(G_2\),\(G_3\))
        \Else \textbf{classify} using \(G_2\)    
        \EndIf
    \Else
        \If{Prediction Probability of \(G_3\) < \(\lambda\)}
            \State Evaluate using \(G_1\), \(G_2\)
            \State MaxProbability(\(G_1\),\(G_2\),\(G_3\))
        \Else \textbf{classify} using \(G_3\)    
        \EndIf
    \EndIf
\end{algorithmic}
\end{algorithm}

\vspace{-0.4cm}
\subsection{Feature Extraction and Model Training} %
\label{sub:feature_extraction}
\vspace{-0.2cm}
After the key clustering step, we then investigate acoustic based features which could be used to identify individual keys. One such set of features include the \emph{Mel Frequency Cepstral Coefficients (MFCC)}, which have been widely used in the field of speech and audio signal processing, particularly for applications such as speech recognition and speaker identification~\cite{ali2020mel}. However, more recently MFCC based features have been used in other acoustic related applications such as keystroke recognition and acoustic activity recognition~\cite{liu2015snooping,6613015}.
The process broadly involves the following steps: (i) First the signal is divided fixed sized frames and for each frame \emph{Fast Fourier Transform (FFT)} is applied to calculate the power spectrum.
(ii) Then, \emph{Mel Filter Bank}
is applied on the power spectrum computed for each frame. The Mel Filter bank is a set of 20-40 (usually) triangular filters that are spaced according to the Mel scale, which approximates the human ear's response more closely than the linearly-spaced frequency bands. This process converts the frequency power spectrum into Mel spectrum \cite{liu2015snooping}.
For our inference framework, we extracted 14 MFCC coefficients for each audio channel (left and right). To capture the variability and characteristics of these coefficients, we computed several statistical measures: mean, standard deviation, skewness, maximum value, median, and minimum value. This resulted in 14×6=84 features for each channel. Thus, combining both channels, we derived a total of 168 features. In addition to the MFCC features, we also included the \emph{Root Mean Square Energy (RMSE)}~\cite{7950204} of each keystroke per channel, bringing the total feature count to 170. 
Building on this, we tested several models including \emph{Random Forest classifier}, \emph{Decision Tree Classifier} and a \emph{Deep Neural Network} for our analysis. 
To optimize its performance, we utilized a Grid Search Cross-Validation approach for hyperparameter tuning. As described in \cref{ssub:energy_based_clustering}, we train three different models for key groups $G_1$, $G_2$ and $G_3$ using labeled training data.
The training and testing were executed in a Leave-One-Out Cross-Validation (LOOCV) manner, ensuring that a test/victim participant was excluded in each iteration.

\vspace{-0.2cm}
\subsection{Keyboard Type Inference} %
\label{sub:keyboard_type_inference}
\vspace{-0.2cm}
A preliminary step for an attacker aiming to execute a keystroke inference attack is to infer the type of keyboard the victim employs. Given the distinct acoustic signatures produced by different keyboard types, such as mechanical or membrane, understanding the keyboard type can pave the way for a more targeted and effective attack. In our study, we gathered data from two distinct brands for each of the key board categories: \textbf{K1}: mechanical, \textbf{K2}: membrane, and \textbf{K3}: laptop-based membrane keyboards. The rationale behind using two models for each category was to introduce a level of complexity to the inference task. If the model were trained solely on data from a single brand for each category, it would trivially classify that brand during testing. Our objective with keyboard type inference is to generalize across brands and variations within each category, ensuring the model can identify the overarching category to which they belong.
For the purpose of type inference, we segment the keyboard input audio data into 30-second windows, extracting 6 MFCC features and Root-Mean-Square-Energy RMSE (\cref{eq:1_energy} for each segment.
We observed during our experiments that, 
while keystroke inference demands a much more detailed feature set, keyboard type inference can be effectively achieved with just these 6 MFCC features. 
Subsequently, we employ a \emph{multi-class logistic regression} model trained on this data to predict the keyboard type. 
The keyboards tested in our keyboard type inference experiment are as follows:
\vspace{-0.2cm}
\begin{itemize}[label={},leftmargin=*]
\setlength{\itemsep}{0pt}
\item
\textbf{K1}: Monoprice MP810 (with red switches) \cite{monoprice_mp810} and Aukey KMG12 (with blue switches) \cite{aukey_kmg12}.
\vspace{-0.18cm}
\item
\textbf{K2}: Logitech K120 \cite{logi_k120} and Dynex DX-PNC2019 \cite{dynex_dxpnc}.
\vspace{-0.18cm}
\item
\textbf{K3}: Tecknet Ultra Slim Compact \cite{tecknet_slim} and the keyboard of the HP Envy x360 15" laptop \cite{hp_envy}.
\end{itemize}
\vspace{-0.2cm}

\vspace{-0.2cm}
\subsection{Word Prediction} %
\label{sub:word_prediction}
\vspace{-0.2cm}
After predicting individual keystrokes, we further look into the possibility of increasing the effectiveness of our attack in a context aware manner by predicting the possible word (comprising of the inferred keystrokes). To this end, we mainly explored two methods to tackle our word prediction task. The first approach is a naive dictionary-based method, where each sequence derived from the top-$k$ predicted letters (from the aforementioned key prediction models) is cross-referenced with a predefined dictionary. If a sequence matches an entry within the dictionary, it's deemed a valid word.  However, this method has its limitations, especially when the key predictions contain inaccuracies such incorrect key predictions, missing certain keystrokes or having extra keystrokes.
Our second method leverages the \emph{SymSpell} algorithm \cite{Garbe_SymSpell_2012}. SymSpell is a spelling correction algorithm that works by pre-computing possible spelling errors for every word in its dictionary, up to a given edit distance. Instead of searching for possible corrections during the lookup, it directly utilizes this pre-computed data to identify close matches. This design allows for rapid and memory-efficient word predictions and corrections, making it particularly suitable for our scenario due to the possibility of presence of missing, incorrect or extra keystrokes.
The procedure (see \cref{alg:spelling_correction}) starts by initializing the SymSpell library and loading a comprehensive frequency dictionary. Next, given a set of top-$k$ letter predictions, we generate all possible word combinations. For each of these generated terms, we consult SymSpell to gather the closest matching words in the dictionary. This results in a collection of suggested words, each associated with its frequency of usage. Finally, to provide the most probable corrections, we sort the accumulated suggestions based on their word frequencies and return the top-$w$ predictions as the output. The rationale being that words that appear more frequently in the language (or specific corpus) are more likely to be the intended word when a spelling error is made. In essence, the frequency of usage helps in prioritizing common words over less common ones when suggesting corrections. 

\begin{algorithm}
\small
\caption{Word Prediction with SymSpell.}
\label{alg:spelling_correction}
\begin{algorithmic}[1]
\Require $predictions$:top-k letter predictions
\Require $topK$: number of words to return

\Function{SpellCorrection}{$predictions$, $topK$}
    \State Initialize $sym\_spell$ and load dictionary
    \State $possible\_combinations \gets$ \Call{GenerateCombinations}{$predictions$}
    \State Initialize $predicted\_words\_with\_counts$ as empty list

    \For{each $input\_term$ in $possible\_combinations$}
        \State Get $suggestions$ for $input\_term$ from $sym\_spell$
        \State Append unique $suggestions$ to $predicted\_words\_with\_counts$
    \EndFor

    \State Sort $predicted\_words\_with\_counts$ by word frequency
    \Return First $top-w$ words from $predicted\_words\_with\_counts$
\EndFunction

\end{algorithmic}
\end{algorithm}

\vspace{-0.2cm}
\subsection{Experimental Setup}
\label{sub:experimental_setup}
\begin{figure}[b]
\centering
\vspace{-0.25cm}
\includegraphics[width=0.5\linewidth]{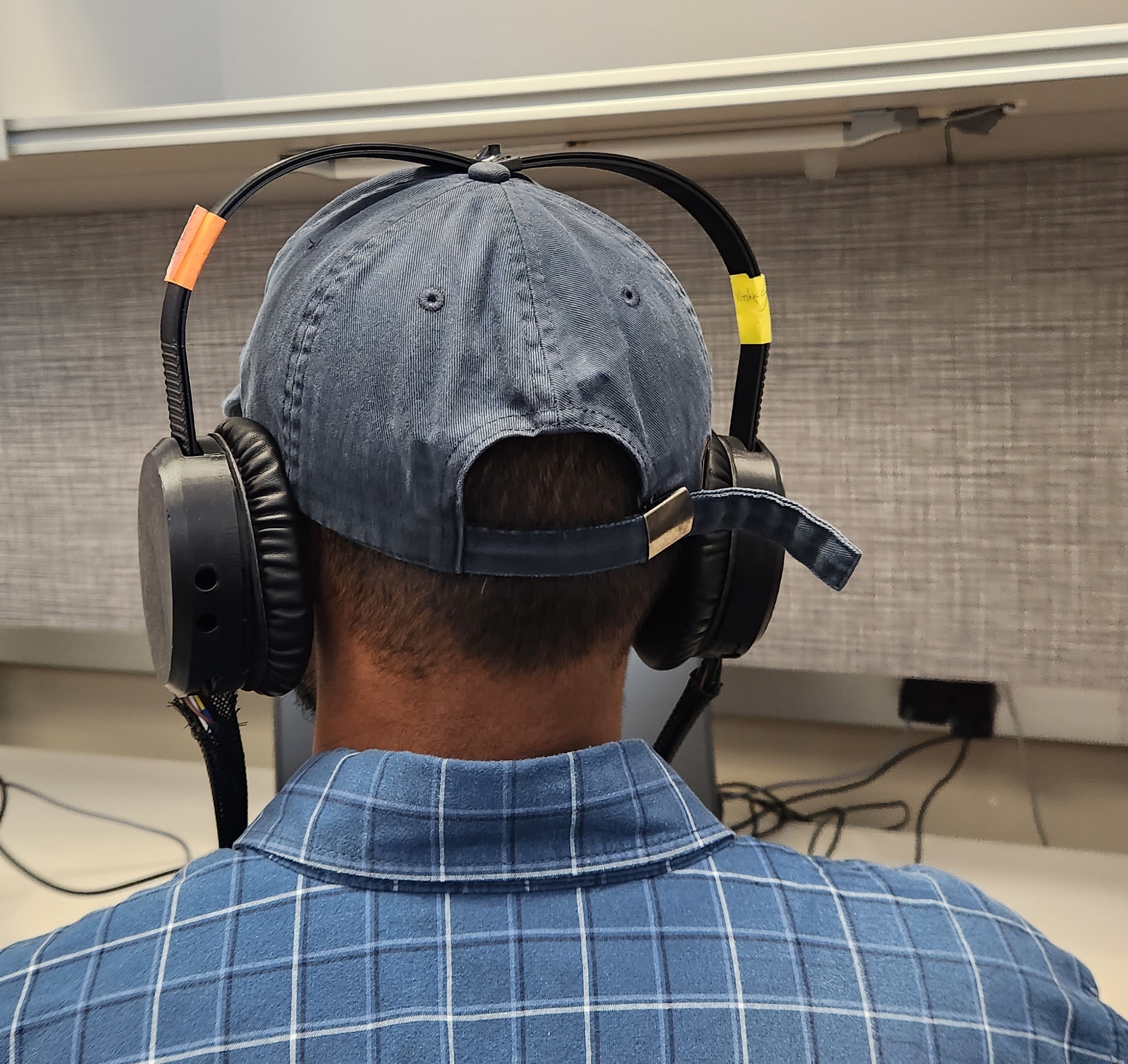}
\caption{A user wearing our headphone setup.}
\label{fig:headphone_setup}
\end{figure}

\vspace{-0.2cm}
Due to the absence of published APIs in current generation of commercial headphones that allow for accelerometer data extraction, we were compelled to devise our own custom setup to evaluate \name. %
Our experimental setup comprises of a Raspberry Pi, and a 3D-printed over-the-ear headphone prototype (see \cref{fig:headphone_setup}). To capture audio, we equipped each earpiece with an Adafruit I2S MEMS Microphone \cite{adafruit_i2s} and a MPU-6500 accelerometer \cite{mpu_6500} to record the motion data. The audio was sampled at a maximum of 96 kHz while accelerometer was sampled at a maximum of 500 Hz.
The microphones and the accelerometers were connected to the Raspberry Pi via the GPIO interface and a Python script was used to record the data from each microphone/sensor. The data processing and inference framework evaluation was done on a Ubuntu 22.04 virtual machine with 32GB memory and 32 cores using Python 3.10.

Our experiments included a diverse range of keyboards to ensure a comprehensive evaluation. We used a AUKEY KMG12 \cite{aukey_kmg12}, a full-sized mechanical keyboard (104 keys) to represent \textbf{K1} category. For the \textbf{K2} category, we utilized a Logitech K120 \cite{logi_k120}, another full-sized model. To closely replicate a keyboard of modern laptops a Tecknet Ultra Slim Compact keyboard (68 keys)\cite{tecknet_slim}, was used (representing the \textbf{K3} category) 
This diversity in keyboard types allowed us to assess the robustness and adaptability of our framework across different tactile feedback mechanisms and form factors.

\vspace{-0.2cm}
\subsection{Data Collection} %
\label{sub:data_collection}
\vspace{-0.2cm}
We recruited 17 participants, in the age range from 18-38, to collect typing data using our custom prototype pair of headphones. The participants conducted two experiments across three types of keyboard while wearing our custom headset (as described in \cref{sub:experimental_setup}. The experiments were conducted inside a closed office space.
For the first experiment, they typed individual keys/English alphabets displayed on the screen one at a time, where each key is repeated for five times at random. The second experiment involved them typing 300 English words from a 5000 most frequent words with number of letters ranging from three to seven \cite{WordFrequency2023}. 
Participants engaged in the typing tasks using a 24-inch computer monitor. To ensure comfort and a natural typing posture, they were provided with a height adjustable chair, allowing them to choose a seating level they found most comfortable. We did not impose any specific typing techniques on the participants; instead, they were encouraged to type in a manner consistent with their daily habits.
The participants also filled out a short survey (see \cref{app:survey_questions})
at the end of the experiments which included questions relating to their typing behaviors and headphone usage. Some useful/interesting insights obtained via the survey are given in \cref{app:survey_responses}.
All participant recruitment and data collection experiments for our study were done under approval from our institution's Institutional Review Board (IRB).

\vspace{-0.2cm}
\section{Evaluation} %
\label{sec:evaluation}
\vspace{-0.35cm}
In this section, we comprehensively evaluate the performance of our proposed keystroke inference framework, \name, 
under a wide variety of different experimental settings and conditions.

\vspace{-0.2cm}
\subsection{Metrics} %
\label{sub:metrics}
\vspace{-0.2cm}
We use the following metrics for quantifying the performance of \name.

\noindent
\textbf{Precision and Recall.} Precision measures the number of correctly predicted keystroke segments out of the total predicted keystroke segments, while recall (or sensitivity) calculates the number of correctly predicted keystroke segments out of the actual keystroke segments. We also use precision and recall to measure the prediction performance 
of our keyboard type inference module of the \name framework.

\noindent
\textbf{Top-$\bm{k}_{key}$ Accuracy.} This metric evaluates the accuracy of the top-$k$ key predictions. Specifically, if the true label is within the top-$k$ predicted labels, then the prediction is considered correct. We utilized top-$k_{key}$ accuracy for assessing the performance of our \name framework at an individual key prediction level.

\noindent
\textbf{Top-$\bm{k}_{word}$ Accuracy.} Evaluates the accuracy of the top-$k$ word predictions. If the true word is within the top-$k$ predicted words, the prediction is deemed accurate. %

\vspace{-0.2cm}
\subsection{Keyboard Type Inference} %
\label{sub:keyboard_type_inference_results}
\vspace{-0.2cm}
Across all keyboard type categories, our keyboard type identification model demonstrated robust performance, consistently achieving an accuracy exceeding 0.95 when the training data includes data from the same brand of keyboard. However, when the type inference model is trained using one keyboard brand for each category, the performance slightly degrades, yet except for K2 category of keyboards, both the other categories (see \cref{tab:kb_type}) demonstrated a recall over 0.95.

\begin{table}[t]
\centering
\small
\caption{Keyboard Type Inference Performance.}
\vspace{-0.275cm}
\label{tab:kb_type}
\begin{tblr}{
cells = {c},
hline{1,5} = {-}{0.08em},
hline{2} = {-}{},
}
\textbf{Keyboard Type} & \textbf{Precision} & \textbf{Recall}\\
Mechanical (\textbf{K1}) & 0.86 & 0.96 \\
Membrane Type 1 (\textbf{K2}) & 0.98 & 0.76 \\
Membrane Type 2 (\textbf{K3}) & 0.88 & 0.99
\end{tblr}
\vspace{-0.2cm}
\end{table}

\vspace{-0.2cm}
\subsection{Keystroke Detection} %
\begin{figure}[H]
\centering
\includegraphics[width=0.85\linewidth]{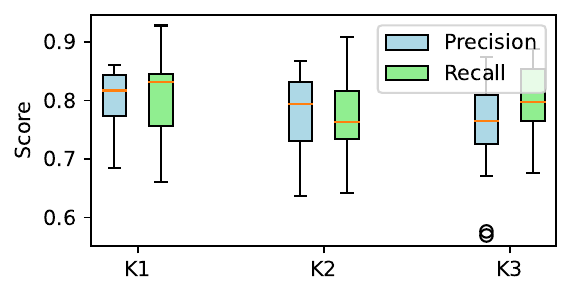}
\vspace{-0.375cm}
\caption{Keystroke detection performance for different keyboards types.}
\label{fig:keystroke_detection}
\vspace{-0.25cm}
\end{figure}

\label{sub:keystroke_detection}
\vspace{-0.2cm}
Our keystroke segmentation algorithm, as part of the \name framework, exhibited consistent performance across various keyboard types (see \cref{fig:keystroke_detection}). For keyboard type \textbf{K1}, the precision and recall were both measured at 0.80 ($\sigma$=0.05 and $\sigma$=0.08, respectively). For keyboard type \textbf{K2}, the precision was 0.78 ($\sigma$=0.07) and the recall was 0.77 ($\sigma$=0.07). For keyboard type \textbf{K3}, we observed a precision of 0.75 ($\sigma$=0.08) and a recall of 0.80 ($\sigma$=0.06).
While these results indicate stability in performance, there are inherent challenges that contribute to the slightly lower accuracy. One primary challenge arises when keys are pressed quickly in succession. Particularly with adept typists, the acoustic energy from one key can overlap with the subsequent key, complicating the distinction between individual keystrokes. Additionally, the unique typing dynamics of each individual introduce variability. Some users exert varied force on keys, while others occasionally press two keys nearly concurrently. These issues add layers of complexity to the keystroke detection process.

\vspace{-0.25cm}

\vspace{-0.2cm}
\subsection{Overall Performance} %
\label{sub:overall_performance}
\vspace{-0.2cm}
As mentioned earlier, we evaluated \name 's performance %
across three keyboard type categories, \textbf{K1}, \textbf{K2}, \textbf{K3}.  The results revealed that, \textbf{K1}, achieved the highest top-5$_{key}$ accuracy of 0.77 and a top-10$_{key}$ accuracy of 0.88.
In comparison, \textbf{K2}, recorded a top-5$_{key}$ accuracy of 0.58, and \textbf{K3}, had a top-5$_{key}$ accuracy of 0.53.
Mechanical keyboards, i.e., category \textbf{K1} tend to produce near distinct tactile feedback and sound profiles for each key press. This unique acoustic signature for each key can make it easier for the system to differentiate between keystrokes, leading to higher accuracy as observed in our results. While larger external membrane keyboards (category \textbf{K2}) too produce a some amount of tactile feedback, the sound profiles might not be as distinct as those of mechanical keyboards. The smaller, membrane type \textbf{K3} which closely resembles laptop keyboards, typically have keys closer together, leading to overlapping or less distinct sound profiles, especially when keys are pressed in rapid succession. Additionally, the build and material of such keyboards might dampen the sound further, making it harder to infer keystrokes accurately.

In evaluating the efficacy of our clustering algorithm, we compared the accuracy of our model with and without the clustering approach. As presented in \Cref{tab:clustering}, the clustering algorithm considerably enhanced the accuracy across all keyboard types. For the mechanical keyboard type, \textbf{K1}, the top-5$_{key}$ accuracy improved from 0.59 to 0.77. Similarly, for the membrane type, \textbf{K2}, there was a noticeable increase from 0.41 to 0.58. The membrane type, \textbf{K3}, also saw an enhancement in accuracy, with top-5$_{key}$ accuracy rising from \(0.37\) to \(0.52\).
These results shows the effectiveness of the accelerometer based clustering algorithm in refining the keystroke inference, making it an important component of our \name inference framework.

\vspace{-0.25cm}
\begin{figure}[h]
\centering
\includegraphics[width=0.8\linewidth]{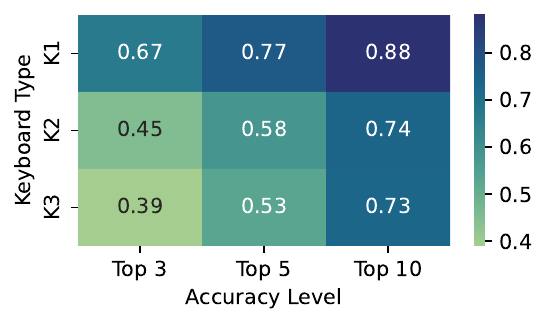}
\vspace{-0.375cm}
\caption{Top-$k_{key}$ accuracy comparison across different keyboard types.}
\label{fig:keyboard_types_acc}
\vspace{-0.25cm}
\end{figure}

\begin{table}[h]
\centering
\small
\vspace{-0.25cm}
\caption{Accuracy comparison (top-5$_{key}$), with and with out the clustering algorithm.}
\vspace{-0.275cm}
\label{tab:clustering}
\begin{tblr}{
cells = {c},
hline{1,5} = {-}{0.08em},
hline{2} = {-}{},
}
\textbf{Keyboard Type} & \textbf{w/o Clustering} & \textbf{with Clustering}\\
\textbf{K1} & 0.59 & 0.77 \\
\textbf{K2} & 0.41 & 0.58 \\
\textbf{K3} & 0.37 & 0.53
\end{tblr}
\end{table}

\begin{figure}[h]
\centering
\includegraphics[width=0.85\linewidth]{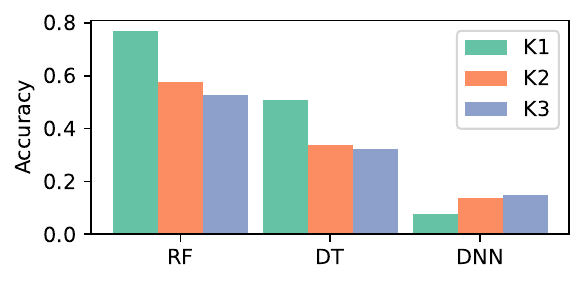}
\vspace{-0.375cm}
\caption{Model comparison (top-5$_{key}$ accuracy).}
\label{fig:model_comparison}
\vspace{-0.25cm}
\end{figure}

In our evaluations, the Random Forest classifier consistently outperformed other models across all keyboard types (see \cref{fig:model_comparison}). Specifically, for \textbf{K1}, it achieved a top-5$_{key}$ accuracy of approximately 0.77, while for \textbf{K2} and \textbf{K3}, the accuracies were 0.57 and 0.53, respectively. In contrast, the Decision Tree classifier managed a top-5$_{key}$ accuracy of 0.57 for \textbf{K1} and around 0.33 for both \textbf{K2} and \textbf{K3}. The Deep Neural Network (DNN) model was the least effective, with accuracies falling below 0.15 for all keyboard categories. The Random Forest classifier outperformed other models likely due to its ensemble nature, effectively capturing complex patterns without overfitting. In contrast, the Deep Neural Network (DNN) model struggled, likely because DNNs require substantial amount of data for effective training, beyond the dataset that we collected. Moving forward, exploring alternative models such as single- and few-shot learning techniques could potentially offer more robust solutions, especially when training data is limited \cite{wang2020generalizing}.

\vspace{-0.2cm}
\subsection{Sampling Rates} %
\begin{figure}[H]
\centering
\includegraphics[width=0.85\linewidth]{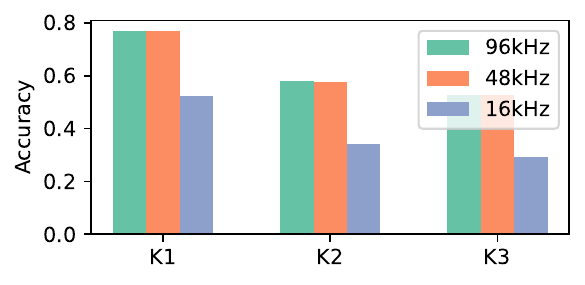}
\vspace{-0.375cm}
\caption{Top-5$_{key}$ prediction accuracy vs. sampling rates.}
\label{fig:topk_samp_rate}
\vspace{-0.25cm}
\end{figure}

\label{sub:sampling_rates}
\vspace{-0.2cm}
In our experiments, we explored the impact of different sampling rates on the performance of our \name inference framework (see \cref{fig:topk_samp_rate}), starting from our default rate of 96kHz for the audio signal. Our findings indicate that the performance at 48kHz is nearly on par with that at 96kHz. However, when the sampling rate is further reduced to 16kHz, we observed a considerable degradation in \name 's performance. This indicates \name can work fairly well even at lower sampling rates.

\vspace{-0.25cm}

\vspace{-0.2cm}
\subsection{Ambient Noise} %
\begin{figure}[h]
\centering
\includegraphics[width=0.85\linewidth]{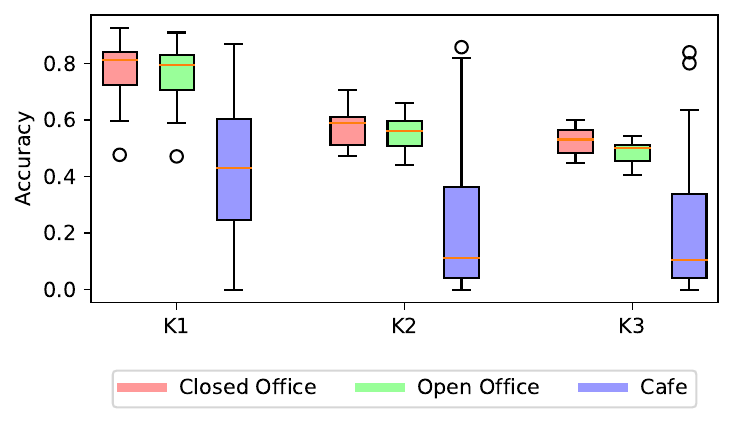}
\vspace{-0.375cm}
\caption{Top-5$_{key}$ accuracy under vs. types of ambient noise.}
\label{fig:noise_types}
\vspace{-0.25cm}
\end{figure}

\label{sub:ambient_noise}
\vspace{-0.2cm}
We evaluated the robustness of \name under various ambient noise conditions. In a university cafeteria setting, the environment was busy with people eating, working on their laptops, and background music playing. The open office space had 2-3 individuals working nearby on computers, accompanied by the typical sounds of typing, mouse clicks, and the occasional mobile phone ringing. In contrast, the closed office space provided a quiet environment with minimal background noise.
From the results (see \cref{fig:noise_types}), it's evident that \name 's accuracy is highest in quieter environments, such as a closed office, and decreases with increasing ambient noise, with the cafeteria setting being the most challenging with accuracies dropping below 0.45. This trend is consistent across all three keyboard types. This shows that \name works well in quieter (ideal) environments, but the accuracy is reasonable even in noisier (less ideal) environments.

\vspace{-0.2cm}
\subsection{Word Prediction} %
\begin{figure}[h]
\centering
\includegraphics[width=0.85\linewidth]{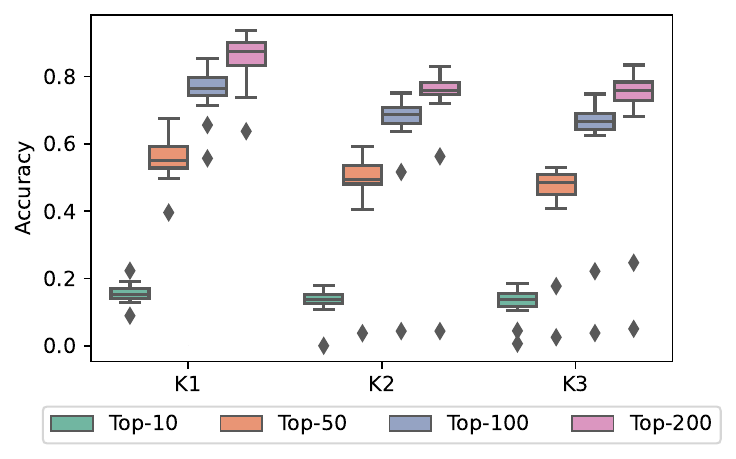}
\vspace{-0.375cm}
\caption{Top-$k_{word}$ accuracies.}
\label{fig:keyboard_types_acc_word}
\vspace{-0.25cm}
\end{figure}

\label{sub:word_prediction_results}
\vspace{-0.2cm}
Our word prediction technique, leveraging the \emph{SymSpell} library, as seen in \cref{fig:keyboard_types_acc_word}, demonstrated considerable efficacy, achieving top-50$_{word}$ accuracies nearing 0.50 across all keyboard categories. Further, \textbf{K1} reached a 0.76 accuracy for top-100$_{word}$ predictions, while \textbf{K2} and \textbf{K3} closely followed with 0.71 and 0.70, respectively. 
For \textbf{K1}, five out of six, participants achieved an accuracy exceeding 0.60 for top-50$_{word}$ predictions, with only one participant falling below the 0.40 mark. In the case of \textbf{K2} and \textbf{K3}, barring one outlier in each category, all participants consistently achieved around the 0.50 accuracy level for top-50$_{word}$ predictions.
In contrast, the naive dictionary-based approach, which relies solely on exact matches, lagged behind. Its limited adaptability to variations in keystroke data meant it consistently registered %
accuracies below 0.4 for all keyboard types.
These results can be attributed to SymSpell's ability to efficiently handle typographical errors, which aids in more accurate word suggestions, especially in the presence of potentially inaccurate key predictions along with extra or missing keystrokes. 

\vspace{-0.2cm}
\subsection{Effect of Typing Speeds} %
\begin{figure}[h]
\centering
\includegraphics[width=0.88\linewidth]{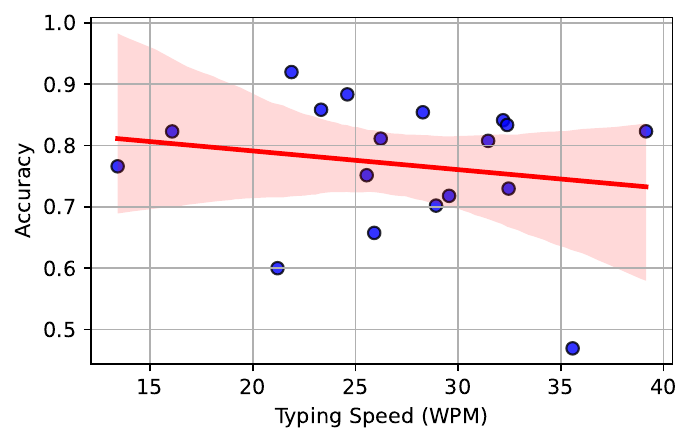}
\vspace{-0.375cm}
\caption{Typing speeds vs. key-level accuracy.}
\label{fig:typing_speeds}
\vspace{-0.25cm}
\end{figure}

\label{sub:effect_of_typing_speeds}
\vspace{-0.2cm}
We initially observed that, faster typing speeds does not impact our keystroke segmentation step (see \cref{app:segmentiation_typing_speed}). Encompassing keystroke segmentation and subsequent processing, \name demonstrates consistent key prediction performance across varying typing speeds (see \cref{fig:typing_speeds}).
While the majority of participants yielded consistent accuracy, an exception was observed in one participant with a typing speed of approximately 35 WPM, who registered an accuracy below 0.5. Despite this outlier, the overall robustness of \name across varied typing speeds is evident, highlighting its adaptability to diverse user behaviors.

\vspace{-0.2cm}
\subsection{Factors Affecting Accuracy} %
\begin{figure}[h]
\centering
\includegraphics[width=0.8\linewidth]{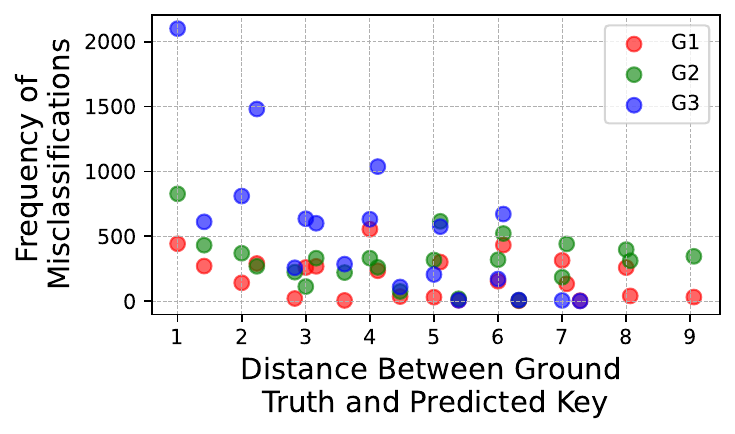}
\vspace{-0.375cm}
\caption{Frequency of misclassifications vs. key distance.}
\label{fig:distance_mis}
\vspace{-0.25cm}
\end{figure}

\label{sub:factors_affecting_accuracy}
\vspace{-0.2cm}
In our analysis of the factors that may be affecting accuracy \name, several patterns emerged.
For the key group $G_2$, no major misclassification patterns were observed. However, there are instances where the key `k' is misinterpreted as `i', and `m' is confused with `n'. In the $G_3$ group, the keys `j' and `h' as well as `y' and `r' are often interchanged. Furthermore, the keys `q' and `w' consistently exhibit lower accuracy rates. In $G_1$ we observed that the key `z' is frequently misclassified as `x', and `x' is often mistaken for `s' (see \cref{app:key_confusions} for further details).
These confusion patterns can be mostly accounted to the spatial closeness of these keys on the keyboard leading to similar acoustic profiles that can be challenging to distinguish.
\Cref{fig:distance_mis} visualizes this relationship between the frequency of misclassifications and the Euclidean distance between ground truth and predicted keys on a QWERTY keyboard, categorized into our three key groups: $G_1$, $G_2$, and $G_3$. A prominent observation from the plot is that misclassifications are more frequent for keys that are closer in distance, particularly for the $G_3$ group. This suggests that keys in the $G_3$ group are often confused with their immediate neighbors on the keyboard. 
The plot shows the inherent challenge in distinguishing between keystrokes of adjacent keys, emphasizing the spatial aspect of the misclassification problem on a physical keyboard layout.

\begin{figure}[h]
\centering
\includegraphics[width=0.8\linewidth]{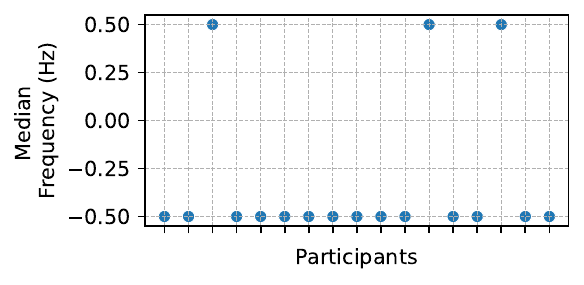}
\vspace{-0.375cm}
\caption{Participants vs. their median gyroscope frequency.}
\label{fig:med_freq_gy}
\vspace{-0.25cm}
\end{figure}

Certain users possess the ability to type without glancing at the keyboard, maintaining a steady gaze on the screen. This consistent posture ensures minimal head movement, leading to relatively stable acoustic profiles. 
Conversely, users who frequently look down at the keyboard introduce regular vertical head movements. These continuous and pronounced shifts can also influence the acoustic signatures, potentially challenging the models employed by \name to accurately distinguish and classify keystrokes. To get a better understanding of such head movements and their potential correlation to the accuracy of \name, we analyzed the frequency spectrum of the gyroscope data (collected alongside accelerometer data with the MPU-6500 sensor used in our custom setup). Particularly, we looked at the median frequency of each participant, which can be considered as the frequency below which 50\% of the power of the signal lies. A higher median frequency in gyroscope data typically indicates more rapid changes in the signal, which can be interpreted as more intense or faster head movements. As it can be seen in \cref{fig:med_freq_gy}, the three participants with the higher median gyroscope showed the lowest top-5$_{key}$ accuracies with values below 0.70 allowing us to potentially hypothesize that more intense head movements might introduce more noise or variability in the audio/accelerometer data, making keystroke inference more challenging.

\vspace{-0.2cm}

\vspace{-0.2cm}
\section{Discussion} %
\label{sec:discussion}
\vspace{-0.35cm}
\textbf{Attack Impact.}
The ubiquity of headphones, equipped with advanced microphones and motion sensors, highlights the potential reach of our keystroke inference attack. Leading products like the Pixel Buds \cite{pixel_buds}, Apple AirPods \cite{airpods}, and the over-ear AirPods Max \cite{airpods_max} not only promise superior audio/voice call quality but also come integrated with accelerometers and gyroscopes for a better user experience. These features, while enhancing user experience, inadvertently make these devices susceptible to \name. Our results, particularly the high accuracy on external mechanical and membrane keyboards, emphasize the feasibility and potential success of such an attack in real-world scenarios. Question 7 of the participant survey (\cref{app:survey_questions}) further corroborates the threat landscape. With 12\% of respondents using mechanical keyboards and 47\% on membrane keyboards it's evident that a large segment of users could be vulnerable.
Specifically, mechanical keyboards have seen a resurgence in popularity over the past decade, especially among certain demographics, such as, the gaming community, technology enthusiasts and professional typists \cite{mech_keyboard_growth}.
Given the projected Compound Annual Growth Rate (CAGR) of 6.79\% from 2022-2027, it's evident that this user base is not only substantial but also expanding \cite{mech_keyboard_growth}. Our \name inference framework, which demonstrates heightened efficacy for mechanical keyboards, poses a notable threat to this growing user base.

\noindent
\textbf{Mitigations.}
Noise-canceling features included in most modern headphones offers a promising avenue to counteract the threat of acoustic keystroke inference. Originally designed to minimize ambient sounds, this can be further optimized to specifically target and suppress the unique acoustic signatures of key presses. While it's challenging to completely mute the sound of keystrokes, integrating such targeted noise-canceling features to the headphones can significantly degrade the quality of captured keystroke sounds, thereby reducing the effectiveness of inference attacks.
Additionally, employing quieter keyboards, such as Scissor keyboards, can further mitigate such attacks. Scissor keyboards \cite{ScissorSwitchKeyboards}, often in laptops, offer a slim profile and quiet typing due to their scissor-like hinge structure. They outperform mechanical keyboards with audible switches and membrane keyboards with rubber domes by minimizing key-press noise. 
Further at an operating system level of the paired smart device (e.g., smartphone or a computer), a system-level service could be introduced to monitor the amount and frequency of data being sent by each application. If an app, such as a headphone companion app, starts offloading unusually large amounts of data or at an unexpected frequency, it can be flagged for review.

\noindent
\textbf{Limitations.}
While \name demonstrates promising results in the context of keystroke inference using headphones, there are several limitations. Given the current accuracy levels, predicting complex passwords, especially those that incorporate a mix of alphanumeric characters, symbols, and varying cases, becomes challenging. This is particularly true for passwords that do not adhere to common linguistic patterns. Further, \name is robust only up to certain levels of ambient noise. In extremely noisy environments, such as noisy cafeterias or situations where multiple overlapping acoustic sources are present, the attack performance is expected to significantly degrade (as also observed by us in \cref{sub:ambient_noise}).

\vspace{-0.2cm}
\section{Conclusion} %
\label{sec:conclusion}
\vspace{-0.35cm}
Headphones have transitioned from mere audio playback devices to sophisticated tools equipped with high-definition microphones and accelerometers. This evolution, while enhancing user experience, has inadvertently opened doors to potential security threats, notably keystroke inference. In this study, we presented \name, a framework that adeptly harnesses both acoustic and accelerometer data from headphones to infer keystrokes achieving a top-5 key accuracy nearing 80\% for mechanical keyboards and 60\% for membrane keyboards. Further, we were able to achieve top-100 word accuracy of over 70\% for all categories of keyboards. While our results highlight the vulnerabilities introduced by modern headphones in real-world scenarios, they also emphasize the importance of understanding and addressing these emerging security challenges.

\bibliographystyle{plain}
\bibliography{references}

\begin{thebibliography}{10}

\bibitem{adafruit_i2s}
Adafruit.
\newblock {Adafruit I2S MEMS Microphone}.
\newblock \url{https://www.adafruit.com/product/3421}, 2023.
\newblock [Online; accessed 15-Oct-2023].

\bibitem{ali2020mel}
Shalbbya Ali, Safdar Tanweer, Syed~Sibtain Khalid, and Naseem Rao.
\newblock Mel frequency cepstral coefficient: a review.
\newblock {\em ICIDSSD}, 2020.

\bibitem{airpods_max}
Apple.
\newblock {AirPods Max}.
\newblock \url{https://www.apple.com/airpods-max/}, 2022.
\newblock [Online; accessed 15-Oct-2023].

\bibitem{airpods}
Apple.
\newblock {AirPods Pro}.
\newblock \url{https://www.apple.com/airpods-pro/specs/}, 2022.
\newblock [Online; accessed 15-Oct-2023].

\bibitem{asonov2004keyboard}
Dmitri Asonov and Rakesh Agrawal.
\newblock Keyboard acoustic emanations.
\newblock In {\em IEEE S\&P}, 2004.

\bibitem{aukey_kmg12}
AUKEY.
\newblock {AUKEY KMG12 Mechanical Keyboard 104key with Gaming Software}.
\newblock
  \url{https://www.aukey.com/products/km-g12-mechanical-keyboard-blue-switches},
  2023.
\newblock [Online; accessed 15-Oct-2023].

\bibitem{bai2021know}
Jia-Xuan Bai, Bin Liu, and Luchuan Song.
\newblock I know your keyboard input: a robust keystroke eavesdropper based-on
  acoustic signals.
\newblock In {\em ACM Multimedia}, 2021.

\bibitem{ring_bbc}
BBC.
\newblock {Ring doorbell 'gives Facebook and Google user data}.
\newblock \url{https://www.bbc.com/news/technology-51281476}, 2020.
\newblock [Online; accessed 15-Oct-2023].

\bibitem{cai2012practicality}
Liang Cai and Hao Chen.
\newblock On the practicality of motion based keystroke inference attack.
\newblock In {\em International Conference on Trust and Trustworthy Computing},
  2012.

\bibitem{chen2015earpieces}
Christina~Summer Chen.
\newblock Earpieces with gesture control, February~5 2015.
\newblock US Patent App. 13/959,109.

\bibitem{cheng2022dictionary}
Kefei Cheng, Wenqi Li, Liang Zhang, Xiangjun Ma, and Jinghao Chen.
\newblock Dictionary attacks based on tdoa using a smartphone.
\newblock In {\em IEEE IAEAC}, 2022.

\bibitem{dynex_dxpnc}
Dynex.
\newblock {Dynex - DX-PNC2019 Wireless Ergonomic Wireless Keyboard}.
\newblock \url{https://www.dynexproducts.com/pdp/DX-PNC2019/6350929}, 2023.
\newblock [Online; accessed 15-Oct-2023].

\bibitem{fan2021headfi}
Xiaoran Fan, Longfei Shangguan, Siddharth Rupavatharam, Yanyong Zhang, Jie
  Xiong, Yunfei Ma, and Richard Howard.
\newblock Headfi: bringing intelligence to all headphones.
\newblock In {\em ACM WiSec}, 2021.

\bibitem{Garbe_SymSpell_2012}
Wolf Garbe.
\newblock {SymSpell}.
\newblock \url{https://github.com/wolfgarbe/SymSpell}, 2012.
\newblock [Online; accessed 15-Oct-2023].

\bibitem{pixel_buds}
Google.
\newblock {Google Pixel Buds requirements and specifications}.
\newblock \url{https://support.google.com/googlepixelbuds/answer/7544332},
  2022.
\newblock [Online; accessed 15-Oct-2023].

\bibitem{hp_envy}
HP.
\newblock {HP ENVY x360}.
\newblock \url{https://support.hp.com/us-en/document/c06038609}, 2023.
\newblock [Online; accessed 15-Oct-2023].

\bibitem{ScissorSwitchKeyboards}
{KBE team}.
\newblock {What Are Scissor Switch Keyboards}.
\newblock \url{https://keyboardsexpert.com/what-are-scissor-switch-keyboards/},
  2023.
\newblock [Online; accessed 15-Oct-2023].

\bibitem{liebich2018signal}
Stefan Liebich, Johannes Fabry, Peter Jax, and Peter Vary.
\newblock Signal processing challenges for active noise cancellation
  headphones.
\newblock In {\em ITG-Symposium: Speech Communication}, 2018.

\bibitem{liu2015snooping}
Jian Liu, Yan Wang, Gorkem Kar, Yingying Chen, Jie Yang, and Marco Gruteser.
\newblock Snooping keystrokes with mm-level audio ranging on a single phone.
\newblock In {\em ACM MobiCom}, 2015.

\bibitem{liu2015good}
Xiangyu Liu, Zhe Zhou, Wenrui Diao, Zhou Li, and Kehuan Zhang.
\newblock When good becomes evil: Keystroke inference with smartwatch.
\newblock In {\em ACM CCS}, 2015.

\bibitem{logi_k120}
Logitech.
\newblock {Keyboard K120}.
\newblock
  \url{https://www.logitech.com/en-us/products/keyboards/k120-usb-standard-computer.920-002478.html},
  2023.
\newblock [Online; accessed 15-Oct-2023].

\bibitem{lu2019keylistener}
Li~Lu, Jiadi Yu, Yingying Chen, Yanmin Zhu, Xiangyu Xu, Guangtao Xue, and
  Minglu Li.
\newblock Keylistener: Inferring keystrokes on qwerty keyboard of touch screen
  through acoustic signals.
\newblock In {\em IEEE INFOCOM}, 2019.

\bibitem{maiti2016smartwatch}
Anindya Maiti, Oscar Armbruster, Murtuza Jadliwala, and Jibo He.
\newblock Smartwatch-based keystroke inference attacks and context-aware
  protection mechanisms.
\newblock In {\em ACM AsiaCCS}, 2016.

\bibitem{mao2012method}
Xiaodong Mao and Noam Rimon.
\newblock Method and apparatus for enhancing the generation of
  three-dimensional sound in headphone devices, April~17 2012.
\newblock US Patent 8,160,265.

\bibitem{marquardt2011sp}
Philip Marquardt, Arunabh Verma, Henry Carter, and Patrick Traynor.
\newblock (sp)iphone: Decoding vibrations from nearby keyboards using mobile
  phone accelerometers.
\newblock In {\em ACM CCS}, 2011.

\bibitem{monoprice_mp810}
Monoprice.
\newblock {Monoprice MP810 Optical Mechanical Gaming Keyboard}.
\newblock \url{www.monoprice.com/product?p_id=34563}, 2023.
\newblock [Online; accessed 15-Oct-2023].

\bibitem{narain2014single}
Sashank Narain, Amirali Sanatinia, and Guevara Noubir.
\newblock Single-stroke language-agnostic keylogging using stereo-microphones
  and domain specific machine learning.
\newblock In {\em ACM WiSec}, 2014.

\bibitem{nieto2015msaf}
Oriol Nieto and Juan~Pablo Bello.
\newblock {Systematic Exploration of Computational Music Structure Research}.
\newblock In {\em International Society for Music Information Retrieval
  Conference}, 2016.

\bibitem{prest2014sports}
Christopher Prest and Quin~C Hoellwarth.
\newblock Sports monitoring system for headphones, earbuds and/or headsets,
  February~18 2014.
\newblock US Patent 8,655,004.

\bibitem{vizio_spy}
Trusted Reviews.
\newblock {Is your smart TV spying on you? All you need to know about smart TVs
  and your privacy}.
\newblock
  \url{https://www.trustedreviews.com/news/smart-tv-privacy-problems-vizio-samsung-lg-sony-panasonic-2952175},
  2020.
\newblock [Online; accessed 15-Oct-2023].

\bibitem{roddiger2022sensing}
Tobias R{\"o}ddiger, Christopher Clarke, Paula Breitling, Tim Schneegans,
  Haibin Zhao, Hans Gellersen, and Michael Beigl.
\newblock Sensing with earables: A systematic literature review and taxonomy of
  phenomena.
\newblock {\em ACM IMWUT}, 6(3), 2022.

\bibitem{6613015}
Joseph Roth, Xiaoming Liu, Arun Ross, and Dimitris Metaxas.
\newblock Biometric authentication via keystroke sound.
\newblock In {\em International Conference on Biometrics (ICB)}, 2013.

\bibitem{sank1984microphones}
Jon~R Sank.
\newblock Microphones.
\newblock In {\em Audio Engineering Society Conference: The Art and Technology
  of Recording}, 1984.

\bibitem{mpu_6500}
TDK.
\newblock {MPU-6500 Six-Axis (Gyro + Accelerometer) MEMS}.
\newblock
  \url{https://invensense.tdk.com/products/motion-tracking/6-axis/mpu-6500/},
  2023.
\newblock [Online; accessed 15-Oct-2023].

\bibitem{mech_keyboard_growth}
technavio.
\newblock {Mechanical Keyboard Market by Distribution Channel, Type, and and
  Geography - Forecast and Analysis 2023-2027}.
\newblock
  \url{https://www.technavio.com/report/mechanical-keyboard-market-industry-analysis},
  2023.
\newblock [Online; accessed 15-Oct-2023].

\bibitem{tecknet_slim}
Tecknet.
\newblock {TECKNET 2.4G Wireless Keyboard, Ultra Slim Compact Computer
  Keyboard}.
\newblock
  \url{https://tecknet.com/products/tecknet-2-4g-wireless-quiet-keyboard},
  2023.
\newblock [Online; accessed 15-Oct-2023].

\bibitem{terlizzi2012systems}
Jeffrey~J Terlizzi.
\newblock Systems and methods for noise cancellation and power management in a
  wireless headset, October~9 2012.
\newblock US Patent 8,285,208.

\bibitem{anker_cam}
TheVerge.
\newblock {Anker finally comes clean about its Eufy security cameras}.
\newblock
  \url{https://www.theverge.com/23573362/anker-eufy-security-camera-answers-encryption},
  2023.
\newblock [Online; accessed 15-Oct-2023].

\bibitem{wang2015mole}
He~Wang, Ted Tsung-Te Lai, and Romit Roy~Choudhury.
\newblock Mole: Motion leaks through smartwatch sensors.
\newblock In {\em ACM MobiCom}, 2015.

\bibitem{7950204}
Jian Wang, Rukhsana Ruby, Lu~Wang, and Kaishun Wu.
\newblock Accurate combined keystrokes detection using acoustic signals.
\newblock In {\em IEEE MSN}, 2016.

\bibitem{wang2020generalizing}
Yaqing Wang, Quanming Yao, James~T Kwok, and Lionel~M Ni.
\newblock Generalizing from a few examples: A survey on few-shot learning.
\newblock {\em ACM Computing Surveys}, 53(3), 2020.

\bibitem{WordFrequency2023}
WordFrequency.
\newblock {Introduction to the top 60,000 words in English}.
\newblock \url{https://www.wordfrequency.info/intro.asp}, 2023.
\newblock [Online; accessed 15-Oct-2023].

\bibitem{xu2012taplogger}
Zhi Xu, Kun Bai, and Sencun Zhu.
\newblock Taplogger: Inferring user inputs on smartphone touchscreens using
  on-board motion sensors.
\newblock In {\em ACM WiSec}, 2012.

\bibitem{yu2021security}
Zhiyuan Yu, Zack Kaplan, Qiben Yan, and Ning Zhang.
\newblock Security and privacy in the emerging cyber-physical world: A survey.
\newblock {\em IEEE Communications Surveys \& Tutorials}, 23(3), 2021.

\bibitem{zhuacoustic2014}
Tong Zhu, Qiang Ma, Shanfeng Zhang, and Yunhao Liu.
\newblock Context-free attacks using keyboard acoustic emanations.
\newblock In {\em ACM CCS}, 2014.

\end{thebibliography}

\appendix
\section{Keystroke Detection vs. Typing Speeds}
\label{app:segmentiation_typing_speed}

\begin{figure}[H]
\centering
\includegraphics[width=0.8\linewidth]{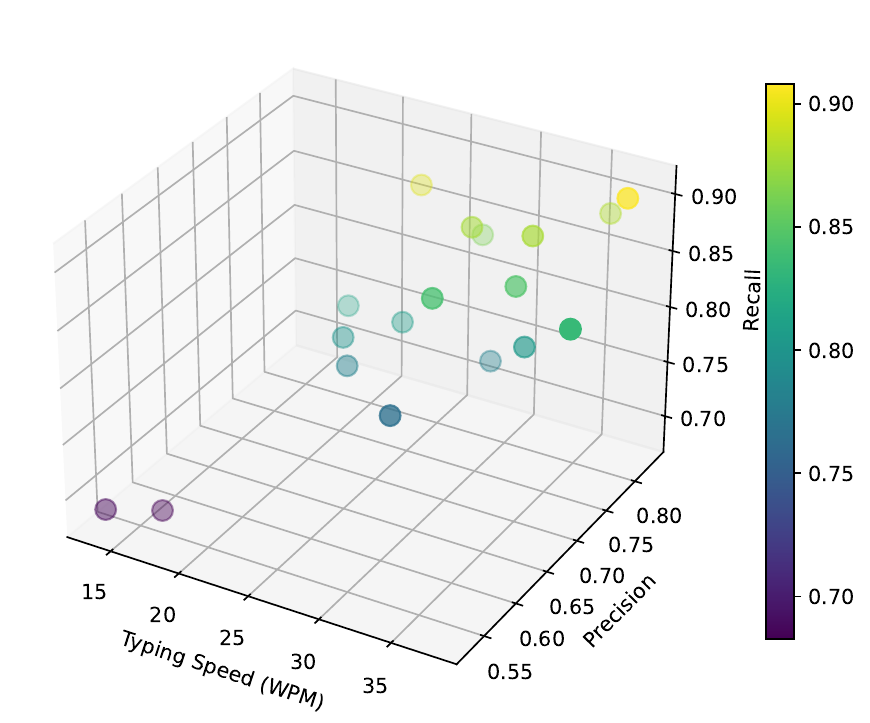}
\caption{Precision and recall for keystroke detection vs. user typing speed (WPM) for \textbf{K1}.}
\label{fig:keystroke_detection_vs_speed}
\end{figure}

\Cref{fig:keystroke_detection_vs_speed} illustrates the precision and recall of our keystroke detection algorithm (detailed in \cref{sub:segmentation} and \cref{sub:keystroke_detection}). Notably, the algorithm's performance remains consistent even at higher typing speeds.

\section{Confusion Matrices for Key Predictions}
\label{app:key_confusions}

\begin{figure}[H]
\centering
\includegraphics[width=0.8\linewidth]{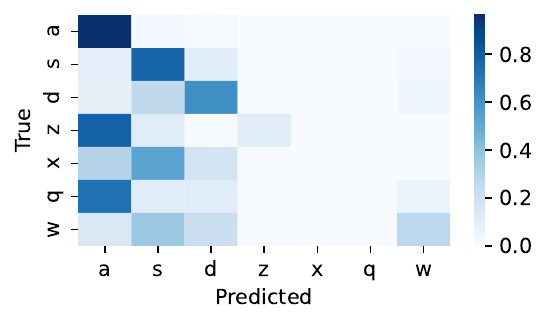}
\caption{Confusion matrix for group $G_1$ keys.}
\label{fig:keys_cm_g1}
\end{figure}

\begin{figure}[h]
\centering
\includegraphics[width=0.8\linewidth]{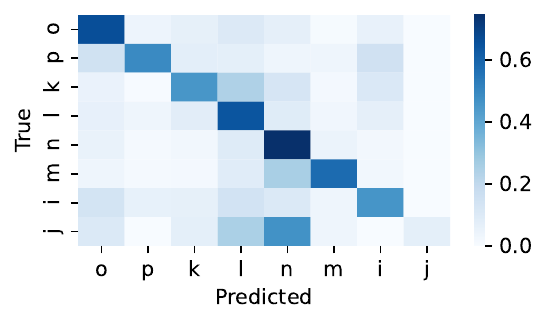}
\vspace{-0.375cm}
\caption{Confusion matrix for group $G_2$ keys.}
\label{fig:keys_cm_g2}
\end{figure}

\begin{figure}[H]
\centering
\includegraphics[width=0.8\linewidth]{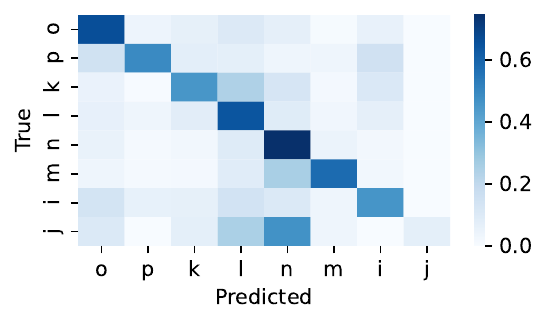}
\caption{Confusion matrix for group $G_3$ keys.}
\label{fig:keys_cm_g3}
\end{figure}

\Cref{fig:keys_cm_g1,fig:keys_cm_g2,fig:keys_cm_g3} present confusion matrices for key predictions across groups $G_1$, $G_2$, and $G_3$. Notably, while $G_2$ and $G_3$ exhibit minimal misclassifications, in $G_1$, most keys surrounding `a' are frequently mistaken for `a'.

\section{Word Prediction Under Noise}
\label{app:noise_word_prediction}

\begin{figure}[H]
\centering
\includegraphics[width=0.85\linewidth]{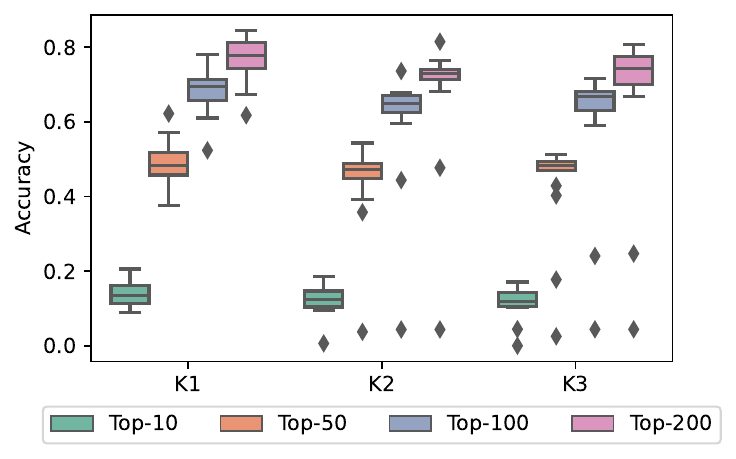}
\caption{Word-level top-w accuracy under cafeteria ambient noise.}
\label{fig:keyboard_types_acc_word_cafe}
\end{figure}

\Cref{fig:keyboard_types_acc_word_cafe} illustrates the word-prediction performance in a cafe ambient noise setting. Despite the key prediction accuracy being notably lower than in quieter environments, top-100$_{word}$ predictions still approached an accuracy close to 0.60.

\section{Participant Survey}
\label{app:survey_questions}
\begin{enumerate}
    \item What is your age?
    \item What is your dominant hand? \\ Right, Left
    \item Do you currently own any of the following digital devices?
    \begin{itemize}
        \item Yes, No - Bluetooth Over-the-Ear Headsets 
        \item Yes, No - Bluetooth Earbuds
    \end{itemize}
    
    \item How often do you type on a computer keyboard while wearing a pair of headsets/earbuds? \\ Never, Rarely, Sometimes, Often, Always
    
    \item How many hours a day do you perform typing tasks in general? 
    
    \item How many hours a day do you perform typing tasks while wearing a headset/earbuds?
    
    \item What type of a keyboard do you own? \\
    Membrane, Mechanical,Not Sure
    
    \item Where do you typically perform your typing tasks? \\
    café, library, classroom, home, other (please specify)
    
    \item Do you keep your headphones/earbuds near your computer/keyboard while not wearing them? \\
    Yes, No
    
    \item Have you installed the smartphone app that comes bundled with your headset/earbuds? \\
    Yes, No
    
    \item Are you aware that modern headphones and earbuds have embedded motion sensors in them? \\
    Yes, No
\end{enumerate}

\section{Insights from Participant Survey Responses}
\label{app:survey_responses}

Through the responses from our participant survey, we were able to gain the below insights into the individual typing behaviors and the prevalent patterns of headphone usage among participants.

\begin{figure}[h]
\centering
\includegraphics[width=0.85\linewidth]{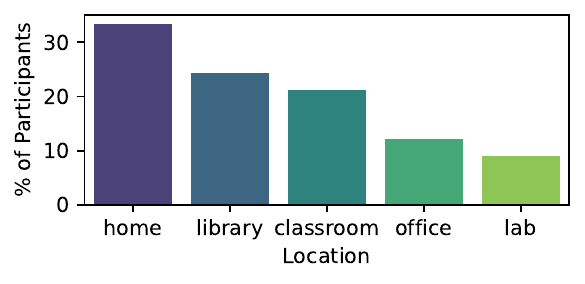}
\caption{Distribution of locations for typing tasks.}
\label{fig:loc_typing_tasks}
\end{figure}

\begin{figure}[h]
\centering
\includegraphics[width=0.85\linewidth]{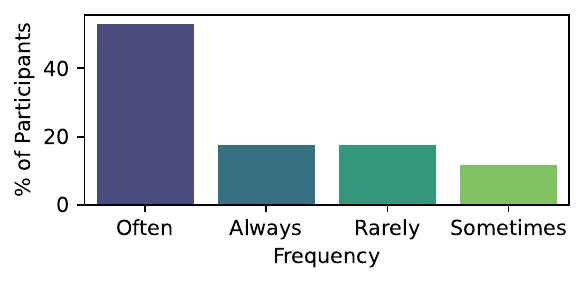}
\caption{Frequency of typing on a computer keyboard while wearing headsets/earbuds.}
\label{fig:typing_while_headset}
\end{figure}

\begin{figure}[h]
\centering
\includegraphics[width=0.85\linewidth]{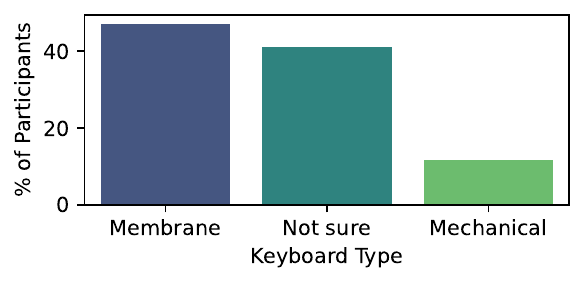}
\caption{Distribution of keyboard types owned.}
\label{fig:keyboard_ownership}
\end{figure}

\begin{figure}[h]
\centering
\includegraphics[width=0.85\linewidth]{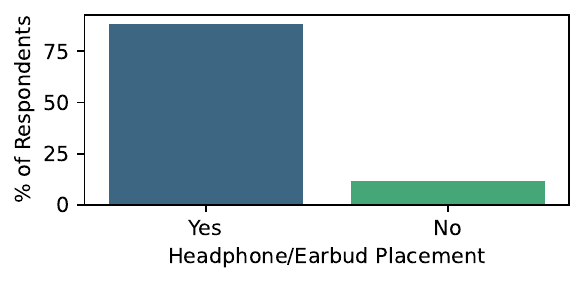}
\caption{Distribution participants who do/do not keep their  headphones/earbuds near the computer.}
\label{fig:headphone_placement}
\end{figure}

\begin{itemize}
    \item A majority of participants (52.94\%) often type on a computer keyboard while wearing headphones or earbuds, with 17.65\% doing so always. On average, 11.76\% type for 2.5 hours per day with these devices on (see \cref{fig:typing_while_headset}).
    
    \item Regarding awareness of modern headphone technology, 52.94\% of participants know that headphones and earbuds often have embedded motion sensors..
    
    \item In terms of keyboard ownership (see \cref{fig:keyboard_ownership}): 
    \begin{itemize}
        \item 47.06\% own a membrane keyboard.
        \item 11.76\% own a mechanical keyboard.
        \item 41.18\% are uncertain about their keyboard type.
    \end{itemize}
    
    \item A significant 88.24\% of participants keep their headphones or earbuds near their computer or keyboard when not in use, while only 11.76\% store them away (see \cref{fig:headphone_placement}).
    
    \item As for typing locations, 33.33\% of participants typically type at their home or apartment, and 24.24\% prefer the library (see \cref{fig:loc_typing_tasks}).
\end{itemize}

\end{document}